\documentclass[english]{article}
\usepackage{geometry}
\usepackage{amsmath}
\usepackage{float} %
\usepackage{xurl}
\usepackage{tabularx}
\usepackage[titletoc]{appendix}
\usepackage{yhmath}
\usepackage{svg}
\usepackage{dsfont}
\usepackage{rotating}
\usepackage[T1]{fontenc}
\usepackage{amsmath,amsfonts,amsthm,bm}
\usepackage{mathtools}
\usepackage{csquotes}
\usepackage{xcolor}
\usepackage{graphicx}
\usepackage{multirow}
\usepackage{amssymb}
\usepackage{wrapfig}
\usepackage[bottom,flushmargin,multiple]{footmisc}
\usepackage{subfigure}
\usepackage{colortbl}
\usepackage{arydshln}
\usepackage{fancyhdr}
\usepackage{wrapfig}
\usepackage{babel}
\usepackage{amsfonts}
\usepackage[font=small,labelfont=bf]{caption}
\usepackage{tabularx}
\usepackage{etoolbox}
\usepackage{adjustbox}
\usepackage{pdflscape}
\usepackage{currency}
\usepackage{lipsum}
\usepackage{enumitem}
\usepackage{wrapfig}
\usepackage{booktabs}
\usepackage{threeparttable}
\usepackage{upgreek}
\usepackage{setspace}
\usepackage[colorlinks=true,urlcolor=blue,citecolor=.,linkcolor=.]{hyperref}
\usepackage{stackengine}
\addto\captionsenglish{}
\addto\captionsenglish{}

\usepackage{scalerel}
\usepackage[backend=biber,style=apa,]{biblatex}
\newcommand{\appendixnumberline}[1]{Appendix\space}

\let\oldappendix\appendix
\makeatletter
\renewcommand{\appendix}{
  \addtocontents{toc}{\let\protect\numberline\protect\appendixnumberline}%
  \renewcommand{\@seccntformat}[1]{Appendix~\csname the##1\endcsname\quad}%
  \oldappendix}

\usepackage{xcolor} 

\title{\huge {Are Higher Suspicious Transaction Report Volumes Associated with More Future Money-Laundering Convictions? } \\ \vspace{0.5cm}
\LARGE Empirical Evidence from the European Union}
\author{
\Large Rasmus Ingemann Tuffveson Jensen\footnote{Corresponding author: \url{rasmus@tuffveson.com}} \\ \large Copenhagen Business School, Denmark \vspace{0.4cm} \\
\Large  Sebastian Holmby Hansen \\ \large University of Copenhagen, Denmark   
\\ The Danish Financial Intelligence Unit, Denmark
\vspace{0.4cm}\\
\Large  Kalle Johannes Rose \\ \large Copenhagen Business School, Denmark  \\}
\date{7 October 2026}

\begin{document}
\doublespacing
\begin{titlepage}
\thispagestyle{fancy}
\fancyfoot{}

    \centering
    \vspace*{2.5cm}

    {\huge \bfseries {Are Higher Suspicious Transaction Report Volumes Associated with More Future Money-Laundering Convictions?
}\par}

    \vspace{1cm}

    {\Large Rasmus Ingemann Tuffveson Jensen\par}
    {\large Copenhagen Business School, Denmark\par}
    {\large \url{rasmus@tuffveson.com}\par}

    \vspace{0.4cm}

    {\Large Sebastian Holmby Hansen\par}
    {\large University of Copenhagen, Denmark\par}
    {\large The Danish Financial Intelligence Unit, Denmark\par}
    {\large \url{seha@di.ku.dk}\par}

    \vspace{0.4cm}

    {\Large Kalle Johannes Rose\par}
    {\large Copenhagen Business School, Denmark\par}
    {\large \url{kjr.acc@cbs.dk}\par}

    \vfill

    {\large 7 October 2026\par}
\end{titlepage}

\setcounter{page}{1}

\begin{abstract}
\noindent Almost all countries require banks to file suspicious transaction reports (STRs) to help authorities detect and prosecute money laundering.  The number of STRs filed in a country is an often-cited metric in international anti-money laundering (AML) compliance evaluations. However, it remains unclear if the volume of STRs filed in a country is associated with the country's future number of money laundering convictions. Using data from Europol, the World Bank, the International Monetary Fund, and the European Sourcebook of Crime and Criminal Justice Statistics, we examine if STR volumes can be associated with future money-laundering conviction volumes in countries in the European Union. Our analysis employs pooled and fixed-effects regression models with one- to five-year lags, controlling for population sizes, shadow-economy estimates, and police force sizes. In our pooled and country-specific fixed-effects models, STR volumes appear to be positively associated with future convictions. However, the association weakens and becomes insignificant when we control for both country-specific and common time effects. Our results provide a caution against assuming that STR volumes, by themselves, are reliable indicators of future money-laundering convictions and, to the extent convictions reflect outcomes, AML effectiveness.
\end{abstract}

\newpage

\section{Introduction}
A meta-analysis published by the United Nations Office on Drugs and Crime \parencite{Pietschmann2001} suggests that money laundering amounts to 2.7\% of the world's combined gross domestic product (GDP). At the same time, scholars have long questioned the effectiveness of global anti-money laundering (AML) efforts \parencite{reuter2004chasing,levi2018can,levi2006,pol2020}. Such efforts are guided by the Financial Action Task Force (FATF), an intergovernmental organization founded in 1989.  Among other things, \textcite{FATF2023} recommends that entities such as banks should be required to report suspicious transactions to national financial intelligence units (FIUs).  The reports are known as suspicious transaction or activity reports (we use the former term)  and are intended to help authorities detect and prosecute money laundering.

In this paper, we examine if the number of suspicious transaction reports (STRs) filed in countries in the European Union (EU) can be associated with the future number of money-laundering convictions in said countries. Following the EU's 4th AML Directive, Member States must have ``effective AML systems'' and obligate relevant entities to file STRs \parencite{AMLD4}. Notably, STR volumes are frequently reported as activity indicators in international AML evaluations \parencite{fatf2022germany,fatf2017sweden}. However, a large volume of STRs does not necessarily indicate an ``effective AML system.'' In particular, the institutional pathway from an STR being filed to it helping secure a criminal conviction (one potential ``effective outcome'') is long: information from the STR must be received and triaged by an FIU, taken up by investigators, and used in a prosecutable case. The quality of STRs, police resources, prosecutorial choices, and court capacity may all affect a potential association between STR volumes and future money-laundering convictions.

\begin{figure}[p]
    \centering    
    \includegraphics[width=0.95\linewidth]{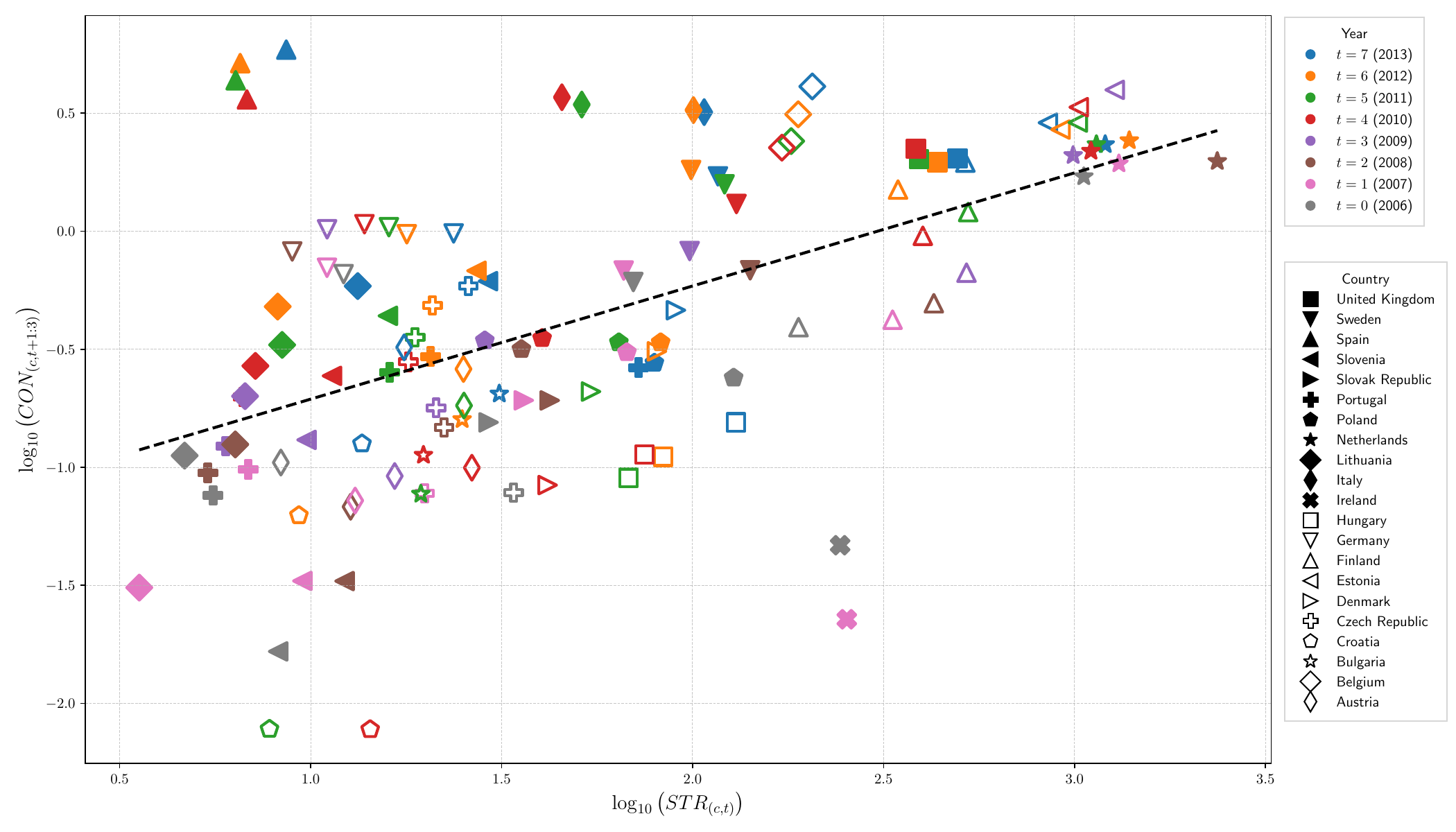}
    \caption{Illustration of our most basic pooled regression model (P1). $STR_{(c,t)}$ denotes the number of STRs filed in country $c$ in year $t$ per 100,000 capita. $CON_{(c,t+1:3)}$ denotes the average number of money-laundering convictions per year in country $c$ up to three years after $t$ per 100,000 capita. We measure $t$ relative to 2006 (the start of our data collection). Thus, $\textit{t}=0$ corresponds to 2006, $\textit{t}=1$ corresponds to 2007, and so on.}
    \label{fig:pooled_reg}
\end{figure}

 Motivated by the observation that different data sources may be inconsistent, we use STR data from a single authoritative source: \textcite{EUROPOL2017}. We stress that this is a practical choice; there are substantial and important differences in STR obligations, thresholds, and AML cultures between different countries (something we discuss in Sections \ref{sec:lit} and \ref{sec:con}).  The data details the number of STRs filed in EU countries between 2006 and 2014. We couple the data with statistics and estimates from the World Bank, the International Monetary Fund, and the European Sourcebook of Crime and Criminal Justice Statistics. The result is an imbalanced panel dataset with two dimensions: country and year. To model the data, we use a log-log specification inspired by the Cobb-Douglas production function \parencite{Cobb1928} and fit pooled and fixed-effects regression models. We stress that our different models examine different questions; our pooled models examine if countries and years with more STRs (\textit{prima facie}) can be associated with higher future conviction volumes; our country-specific fixed-effects models examine if within-country changes in STR volumes can be associated with future conviction volumes; finally, our models with both country-specific and time effects examine if residual deviations left after controls for both country-specific and common time dynamics can be associated with future conviction volumes. For illustration purposes, Figure \ref{fig:pooled_reg} depicts our most basic pooled regression model.

Initial results from our pooled models indicate that countries and years with more STRs can be associated with more future money-laundering convictions, and that the association is sub-linear. The association remains when we add country-specific fixed effects. However, when we include either (i) a linear time trend or (ii) time fixed effects, the relevant coefficient becomes substantially smaller and statistically insignificant. This indicates that STR volumes, at least by themselves, are an insufficient predictor of future convictions once stable country differences and common time effects are removed.

The rest of our paper is organized as follows. Section \ref{sec:lit} reviews related work and discusses the relationship between STRs and convictions. Section \ref{sec:data} presents our data. Section \ref{sec:method} presents our methodology. Section \ref{sec:results} presents our results. Finally, Section \ref{sec:con} contains a conclusion and discussion.

\section{Related Literature} \label{sec:lit}

Scholars have long questioned the effectiveness of global AML efforts and policies \parencite{reuter2004chasing,levi2018can,levi2006,pol2020}. \textcite{levi2006} note that ``how well the [AML] system works in suppressing crimes and preventing terrorist acts is entirely a matter of speculation.'' \textcite{levi2018can} argue that evaluation of AML efforts is hindered by an absence of reliable data, noting that AML evaluations and assessments (including those by FATF) mostly rely on rhetorical arguments and narrative judgments. More directly, \textcite{pol2020,pol2018anti} calls AML ``the world's least effective policy experiment,'' arguing that less than 0.1\% of criminal proceeds are seized while AML costs vastly exceed assets recovered. \textcite{pol2018anti}, in particular, highlights an excessive focus on activity metrics (e.g., the number of STRs filed) over outcomes (e.g., the number of convictions achieved or reduced crimes) in the global AML efforts.

Our study is situated within the broader academic debate on ``AML activity versus effectiveness.'' As noted by \textcite{ferwerda2022} (though in the context of national risk assessments), analyses of aggregate STR counts can often be misleading. In particular, it is not given that STR volumes track the underlying goals of AML policy. A high number of STRs may indicate AML activity, but not necessarily the disruption of money laundering. Indeed, the theoretical relationship between STR and conviction volumes is uncertain; it is not clear that more STRs automatically are associated with more convictions. In the absence of extensive empirical studies, the literature is influenced by a theoretical model from \textcite{takats2011}. The model shows that when financial institutions face significant fines for failing to file ``enough'' STRs, they may respond by filing large volumes of low-value STRs.\footnote{From conversations with AML officers at multiple banks, we are aware that the officers colloquially use the terms ``defensive reports'' or ``cover-my-ass reports'' to describe such STRs, terms we believe capture the essence of the crying-wolf phenomenon.} This behavior leads to an overload of low-quality information being sent to FIUs (a ``crying wolf'' effect), reducing the value of STRs. The model's predictions align with empirical evidence from the United States (US), where STR volumes have increased without an increase in convictions. The model is also backed up by work from \textcite{dalla2023}, finding that ``data suggest that Italian bankers are likely to cry wolf.'' Meanwhile, other work from Italy by \textcite{gara2020looking} could suggest ``substantial under-reporting of high-risk STRs by selected banks.''

We stress that the institutional pathway from an STR being filed until it potentially helps secure a criminal conviction is long: the STR must be triaged by an FIU, taken up by investigators, and used in a prosecutable case. In turn, the effects of STRs may be mediated by institutional factors and country-specific AML regimes. \textcite{unger2009} compare AML regimes in the US and the Netherlands. After introducing a risk-based approach in the 1990s and 2000s, the US saw an increase in reports but a decrease in convictions for money laundering (consistent with the ``crying wolf'' phenomenon). Meanwhile, the Netherlands saw a decrease in reports but an increase in convictions. The authors argue that this is related to different legal traditions; the US being adversarial and the Dutch being cooperative. We note that the EU introduced its 4th AML Directive in 2015, promoting a ``risk-based approach'' to AML \parencite{AMLD4}. While data is scarce and fragmented, the directive does not appear to have caused a general decrease in STRs; see, for example, \textcite{BdI2016,BdI2015,FIUD2016}. Meanwhile, we are not familiar with empirical evidence that the directive produced a more effective (however measured) AML regime. Even after the 4th AML Directive, there are substantial and important differences in STR obligations, thresholds, and AML cultures between different countries. Furthermore, as \textcite{levi2018can} note, there ``is no comparative analysis to show that STRs in countries that apply more stringent criteria (...) are comparable to STRs in low threshold countries (...).'' The organization of FIUs may also affect the utility and utilization of STRs. \textcite{bartolozzi2022} consider factors such as financial and informative powers, investigative authority, independence, and accountability, finding substantial differences across financial intelligence units in different countries. These institutional differences almost certainly affect whether and how STRs (i) are converted into actionable intelligence, (ii) are investigated, and (iii) ultimately may be associated with convictions. Finally, the type of criminal activity and money laundering taking place in a country may naturally affect the utility and utilization of STRs. As noted by \textcite{Pietschmann2001}, it is, for example, likely that ``the geographic concentration of cocaine trafficking activities in the Western hemisphere'' means that the banking sector (typically filing more STRs than other sectors of the economy) plays a larger role in money laundering in the Western hemisphere.

We stress that (new) money laundering convictions are not the only potential ``effective'' outcome of STRs. Indeed, STRs may contribute to ongoing investigations and lead authorities toward  other (predicate) crimes. The point is illustrated by \textcite{nazzari2025possible}, who find that STRs in Italy contribute relevant information to ongoing criminal proceedings and that many STR-prompted proceedings concern offences other than money laundering. We note that some publications, particularly \textcite{EUROPOL2017}, suggest that the ``conversion rate'' of STRs (i.e., the proportion of STRs that lead to active investigations) is a more appropriate measure of STR usefulness. This is certainly understandable from the viewpoint of law enforcement agencies.  At the same time, however, it can be debated if active investigations, by themselves, are outcomes or just another type of activity metric.\footnote{We stress that we do not test if STRs have preventive, disruptive, or intelligence effects not captured by money-laundering convictions. We provide no evidence regarding the existence or non-existence of such effects.}

\section{Data} \label{sec:data}
To build our dataset, we combine publicly available statistics and estimates from Europol, the World Bank, the International Monetary Fund, and the European Sourcebook of Crime and Criminal Justice Statistics. The result is an imbalanced panel dataset with two dimensions: country and year. Our dataset contains information on 28 EU countries between 2006 and 2016. The primary variables in our dataset are:
\renewcommand{\labelitemi}{\hspace*{-0em}}
\begin{itemize}
    \item $STR_{(c,t)}$: the number of STRs filed in country $c$ in year $t$,
    \item $POP_{(c,t)}$: the population size of country $c$ in year $t$,
    \item $GDP_{(c,t)}$: the GDP of country $c$ in year $t$ in constant 2015 USD,
    \item $SHW_{(c,t)}$: the estimated size of the shadow economy of country $c$ in year $t$ as a percentage of GDP,
    \item $POL_{(c,t)}$: the number of police officers employed in country $c$ in year $t$, and
    \item $CON_{(c,t)}$: the number of people convicted for money laundering in country $c$ in year $t$.
\end{itemize}
By multiplying $SHW_{(c,t)}$ and $GDP_{(c,t)}$, we construct another variable:
\begin{itemize}
    \item $SGDP_{(c,t)}$: the estimated size of the shadow economy of country $c$ in year $t$ in constant 2015 USD.
\end{itemize}
The idea behind $SGDP_{(c,t)}$ (shorthand for ``Shadow GDP'') is to express the size of a country's shadow economy in monetary terms rather than as a percentage of GDP (something more natural for the models we consider in Section \ref{sec:method}; see Section \ref{subsec:IMF} for a discussion of the estimates behind $SGDP_{(c,t)}$).  In our models, tables, and graphs, we measure $GDP_{(c,t)}$ and $SGDP_{(c,t)}$ per capita (simply dividing by $POP_{(c,t)}$). Meanwhile, we measure $STR_{(c,t)},POL_{(c,t)},$ and  $CON_{(c,t)}$ per 100,000 capita (dividing by $POP_{(c,t)}$ and multiplying by 100,000). To ease notation later, we measure $t$ relative to 2006 (the start of our data collection). Thus, $\textit{t}=0$ corresponds to 2006, $\textit{t}=1$ corresponds to 2007, and so on.

Table \ref{tab:stats} presents summary statistics for our dataset. Note that there is large variation in $STR_{(c,t)}$, having a standard deviation of 430.23 (substantially more than its mean of 231.47). Table \ref{tab:corr} presents correlation coefficients for our dataset (calculated by comparing observations from the same year, i.e., without a lag). Note that $GDP_{(c,t)},SHW_{(c,t)}$ and $SGDP_{(c,t)}$ (quite naturally) are strongly correlated. While $STR_{(c,t)}$ and $t$ only appear to be weakly correlated, an inspection of our dataset reveals a highly non-linear increase in STRs within countries over time. Indeed, the number of STRs per 100,000 capita increased by almost 240\% on average, taken across all countries, between 2006 and 2014 (with a median increase of approximately 140\%).

The following subsections present each of our data sources and variables in detail. Subsection \ref{subsec:STR Data} presents our $STR_{(c,t)}$ statistics from Europol. Subsection \ref{subsec:World Bank} presents our $GDP_{(c,t)}$ and $POP_{(c,t)}$ statistics from the World Bank.  Subsection \ref{subsec:IMF} presents our $SHW_{(c,t)}$ estimates from the International Monetary Fund. Finally, Subsection \ref{subsec:sourcebook} presents our $POL_{(c,t)}$ and $CON_{(c,t)}$ statistics from the European Sourcebook of Crime and Criminal Justice Statistics.

\begin{table}[p]
    \centering
    \footnotesize
    \rowcolors{2}{white}{gray!25}
    \begin{tabular}{rrrrrrrrr}
\toprule
\multicolumn{1}{c}{} & \multicolumn{1}{c}{$STR_{(c,t)}$} & \multicolumn{1}{c}{$GDP_{(c,t)}$} & 
\multicolumn{1}{c}{$SHW_{(c,t)}$} & \multicolumn{1}{c}{$SGDP_{(c,t)}$} &
\multicolumn{1}{c}{$POL_{(c,t)}$} & \multicolumn{1}{c}{$CON_{(c,t)}$} & 
\multicolumn{1}{c}{$POP_{(c,t)}$} & \multicolumn{1}{c}{$t$} \\
\midrule
Mean & 231.47 & 29,774.35 & 22.98 & 5,624.12 & 343.01 & 0.99 & 18,022,396.17 & 5.00 \\
Std. dev. & 430.23 & 20,825.57 & 8.07 & 2,294.32 & 104.85 & 1.31 & 22,840,247.64 & 3.17 \\
Minimum & 3.56 & 5,629.80 & 9.10 & 2,037.99 & 58.18 & 0.00 & 405,308.00 & 0.00 \\
Median & 41.69 & 21,507.64 & 21.40 & 5,382.46 & 340.16 & 0.34 & 8,908,586.50 & 5.00 \\
Maximum & 2,364.41 & 112,417.88 & 38.50 & 10,618.44 & 570.67 & 6.45 & 82,376,451.00 & 10.00 \\
\bottomrule
\end{tabular}
 
    \caption{Summary statistics for our dataset, including countries and years with missing observations. ``Std. dev.'' denotes standard deviation. We use $t$ to denote the year of an observation, measured relative to 2006. Thus, $\textit{t}=0$ corresponds to 2006, $\textit{t}=1$ corresponds to 2007, and so on. Note that $GDP_{(c,t)}$ and $SGDP_{(c,t)}$ are measured per capita. Meanwhile, $STR_{(c,t)},POL_{(c,t)},$ and $CON_{(c,t)}$ are measured per 100,000 capita.}
    \label{tab:stats}
\end{table}

\begin{table}[p]
    \centering
    \footnotesize
    \rowcolors{2}{white}{gray!25}
    \begin{tabular}{ccccccccc}
\toprule
\multicolumn{1}{c}{} & \multicolumn{1}{c}{$STR_{(c,t)}$} & \multicolumn{1}{c}{$GDP_{(c,t)}$} & 
\multicolumn{1}{c}{$SHW_{(c,t)}$} & \multicolumn{1}{c}{$SGDP_{(c,t)}$} &
\multicolumn{1}{c}{$POL_{(c,t)}$} & \multicolumn{1}{c}{$CON_{(c,t)}$} & 
\multicolumn{1}{c}{$POP_{(c,t)}$} & \multicolumn{1}{c}{$t$} \\
\midrule
$STR_{(c,t)}$ & \phantom{-}1.00 & \phantom{-}0.26 & -0.17 & \phantom{-}0.12 & -0.24 & \phantom{-}0.45 & -0.16 & \phantom{-}0.07 \\
$GDP_{(c,t)}$ & \phantom{-}0.26 & \phantom{-}1.00 & -0.76 & \phantom{-}0.80 & -0.50 & \phantom{-}0.23 & \phantom{-}0.07 & \phantom{-}0.02 \\
$SHW_{(c,t)}$ & -0.17 & -0.76 & \phantom{-}1.00 & -0.43 & \phantom{-}0.38 & -0.10 & -0.28 & -0.02 \\
$SGDP_{(c,t)}$ & \phantom{-}0.12 & \phantom{-}0.80 & -0.43 & \phantom{-}1.00 & -0.30 & \phantom{-}0.35 & \phantom{-}0.03 & \phantom{-}0.01 \\
$POL_{(c,t)}$ & -0.24 & -0.50 & \phantom{-}0.38 & -0.30 & \phantom{-}1.00 & -0.06 & -0.02 & -0.11 \\
$CON_{(c,t)}$ & \phantom{-}0.45 & \phantom{-}0.23 & -0.10 & \phantom{-}0.35 & -0.06 & \phantom{-}1.00 & \phantom{-}0.33 & \phantom{-}0.34 \\
$POP_{(c,t)}$ & -0.16 & \phantom{-}0.07 & -0.28 & \phantom{-}0.03 & -0.02 & \phantom{-}0.33 & \phantom{-}1.00 & \phantom{-}0.01 \\
t & \phantom{-}0.07 & \phantom{-}0.02 & -0.02 & \phantom{-}0.01 & -0.11 & \phantom{-}0.34 & \phantom{-}0.01 & \phantom{-}1.00 \\
\bottomrule
\end{tabular}
 
    \caption{Correlation coefficients (without any time lag) over all countries and years in our dataset. Correlations are computed pairwise using available observations.  We use $t$ to denote the year of an observation, measured relative to 2006. Thus, $\textit{t}=0$ corresponds to 2006, $\textit{t}=1$ corresponds to 2007, and so on.}
    \label{tab:corr}
\end{table}

\subsection{Statistics on Suspicious Transaction Reports From Europol} \label{subsec:STR Data}
Different countries have different laws regarding when and how reports about suspicious transactions should be filed. In turn, the definition of (and name given to) an STR is open to debate, making it difficult to obtain comparable statistics from national authorities. Reports are sometimes denoted as suspicious activity reports (SARs), unusual transaction reports (UTRs), or notifications. The Netherlands, for example, operates a UTR regime \parencite{EUROPOL2017}, Denmark primarily operates an STR regime \parencite{FATF2017DK}, and Poland operates a regime with multiple types of SARs and notifications \parencite{moneyval2022poland}. 

Motivated by the observation that different sources may be inconsistent, we use STR data from a single authoritative source: \textcite{EUROPOL2017}.\footnote{We note that even statistics published in the annual reports of  the same FIU can be inconsistent from year to year (we omit examples in order not to single out individual authors or FIUs). } In turn, we follow \textcite{EUROPOL2017} and define an STR as any report received by a national financial intelligence unit from an obligated entity, whether called an STR, SAR, UTR, or notification. We stress that our choice to use \textcite{EUROPOL2017} as a single source is motivated by practicality. The data compiled by \textcite{EUROPOL2017} was received from national authorities. To our knowledge, \textcite{EUROPOL2017} has not undertaken an effort to address underlying national differences. However, as defined above,  \textcite{EUROPOL2017} lists the number of STRs filed in each EU country from 2006 to 2014, yielding our variable:

\begin{itemize}
\item $STR_{(c,t)}$: the number of STRs filed in country $c$ in year $t$.
\end{itemize}

\noindent Given 28 countries and a data collection period of 9 years, we have 252 observations of $STR_{(c,t)}$. In our models, tables, and graphs, we measure $STR_{(c,t)}$ per 100,000 capita.

\subsection{Statistics on GDP and Population Sizes From the World Bank} \label{subsec:World Bank}
We use the online database of the \textcite{worldbank_wdi} to obtain the following variables:
\begin{itemize}
     \item $GDP_{(c,t)}$: the GDP (World Bank indicator code \texttt{NY.GDP.MKTP.KD}) of country $c$ in year $t$ in constant 2015 USD and

     \item $POP_{(c,t)}$: the population size (World Bank indicator code \texttt{SP.POP.TOTL}) of country $c$ in year $t$.
\end{itemize}

\noindent We collect both variables from 2006 (the start of our STR data) to 2016 (the end of our conviction data). Given 28 countries and a data collection period of 11 years, we have 308 observations of each variable. In our models, tables, and graphs, we measure $GDP_{(c,t)}$ per capita.

\subsection{Estimates of the Shadow Economy from the International Monetary Fund} \label{subsec:IMF}
 We use shadow economy estimates published by the International Monetary Fund \parencite{kelmanson2021}. The estimates build on work by \textcite{kelmanson2019} and are derived from a multiple indicators, multiple causes model. The model infers the size of a country's shadow economy from macroeconomic indicators, including  productivity, tax revenue, trade volume, labor force participation, and investment; see \textcite{kelmanson2021} and \textcite{kelmanson2019}. Thus, we record the following variable from \textcite{kelmanson2021}:
\begin{itemize}
\item $SHW_{(c,t)}$: the estimated size of the shadow economy of country $c$ in year $t$ as a percentage of GDP. 
\end{itemize}

\noindent We collect the variable from 2006 (the start of our STR data) to 2016 (the end of our conviction data). Given 28 countries and a data collection period of 11 years, we have 308 potential observations of $SHW_{(c,t)}$. However, \textcite{kelmanson2021} do not report estimates for Malta. In turn, we only have 297 observations.

 We stress that ``the shadow economy'' and ``money laundering'' are distinct but related concepts. In particular, the shadow economy is defined as economic activity hidden from authorities \parencite{kelmanson2021,kelmanson2019,hassan2016,williams2016}. This will largely include activities that, in and of themselves, are legal. Furthermore, such activities do not necessarily involve money laundering, just as all criminal activities do not require money laundering.\footnote{Imagine, for example, a day laborer who is paid in cash. The laborer's work may, in and of itself, be legal. If the laborer fails to declare their income, it constitutes tax evasion. However, if the laborer simply uses their cash salary (including what should have been paid in taxes) to pay for daily expenses such as groceries, they may not be involved in any kind of money laundering.}

We also stress that $SHW_{(c,t)}$ (and $SGDP_{(c,t)}$, as used by our models) is an estimate from a macroeconomic model, not an observed quantity. In turn, the variable is associated with an uncertainty not properly propagated through to our regression models. In particular, the variable may capture broad structural differences between countries rather than the underlying volume of laundering opportunities. When we, nevertheless, use the variable, it is intended as a rough contextual control. Note that we fit multiple models both with and without the variable; it is only intended for additional robustness.

\subsection{Statistics on Police Forces and Convictions for Money Laundering From the European Sourcebook of Crime and Criminal Justice Statistics} \label{subsec:sourcebook}
We use the European Sourcebook of Crime and Criminal Justice Statistics \parencite{aebi2021,aebi2017,aebi2010} as our source of police and conviction statistics.\footnote{We extract and use raw statistics from the Sourcebook's databases, available on the Sourcebook's website \parencite{aebi2021}. These statistics may at times differ from those published in the print versions of the Sourcebook.} The Sourcebook collects crime and criminal justice statistics from national correspondents. The 6th edition of the Sourcebook, for example, contains statistics from the Danish Ministry of Justice, the Latvian Statistical Bureau, and numerous academics from across Europe. Notably, the Sourcebook aims to collect standardized statistics: a prerequisite for comparing statistics across countries.\footnote{As an alternative data source, we considered using mutual evaluation reports from FATF. However, statistics in these are not standardized. For example, \textcite{FATF2017DK} combines statistics on money laundering and stolen goods. While the Sourcebook aims to collect standardized statistics, these may deviate. In this work, we simply use the statistics as given in the Sourcebook.} The Sourcebook explicitly defines money laundering as ``specific financial transactions to conceal the identity, source, and/or destination of money or non-monetary property deriving from criminal activities.'' We collect the following variables from the Sourcebook:

\begin{itemize}
    \item $POL_{(c,t)}$: the number of police officers employed (Sourcebook indicator code \texttt{T14OP}) in country $c$ in year $t$ and
    \item $CON_{(c,t)}$: the number of people convicted for money laundering (Sourcebook indicator code \texttt{T31ML}) in country $c$ in year $t$.
\end{itemize}

\begin{figure}[p]
    \centering
\includegraphics[width=1\linewidth]{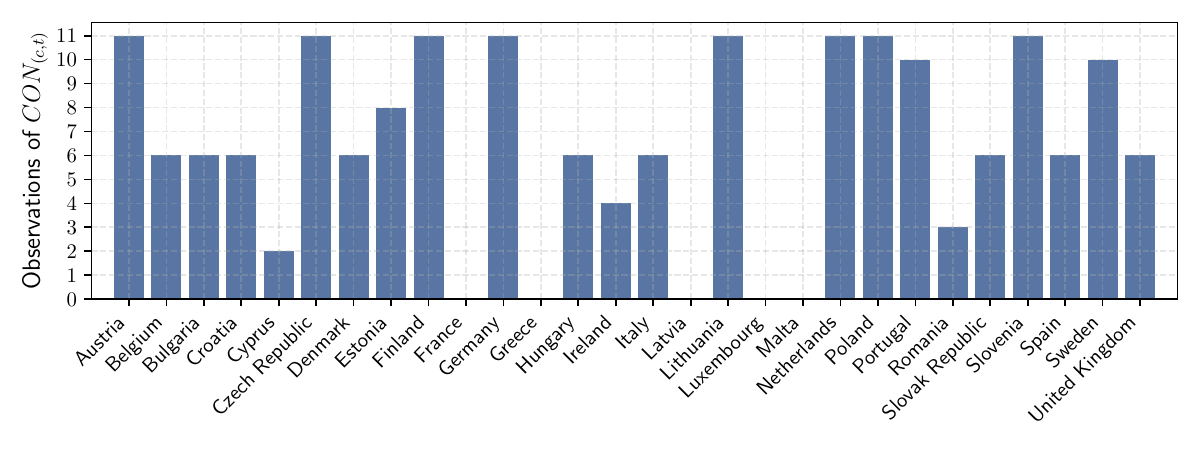}
    \caption{Observations of $CON_{(c,t)}$ per country $c$.}
    \label{fig:country}
\end{figure}

\begin{figure}[p]
    \centering
    \includegraphics[width=1\linewidth]{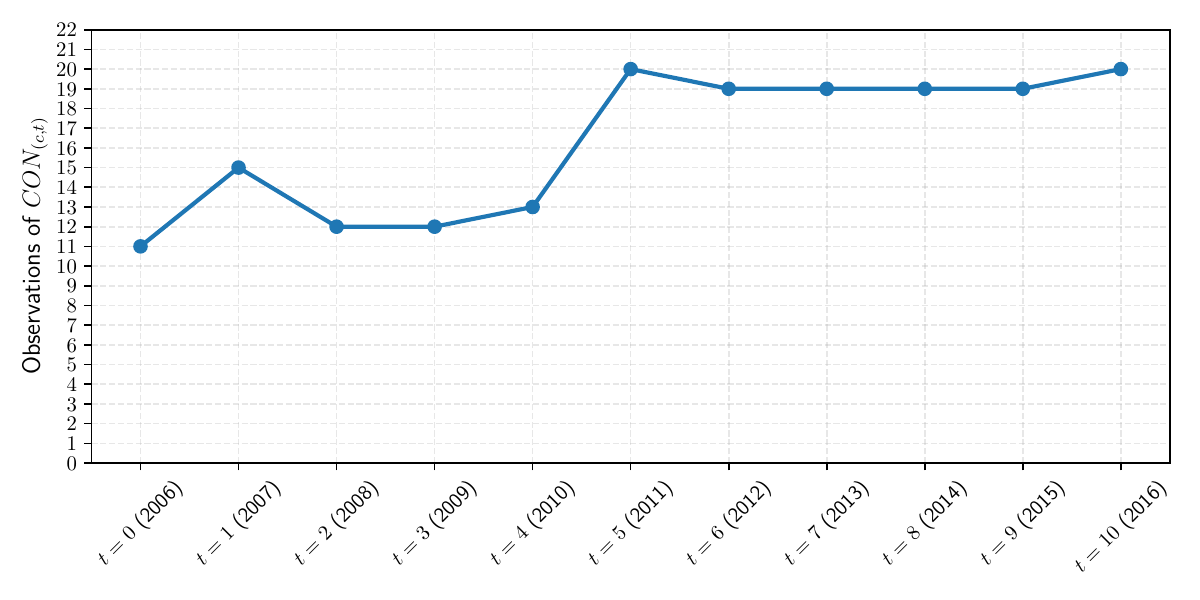}
    \caption{Observations of $CON_{(c,t)}$ per year $t$. Note that $t$ is measured relative to 2006 (the first year of our data).}
    \label{fig:year}
\end{figure}

\noindent We collect both variables from 2006 (the start of our STR data) to 2016  (two years after the end of our STR data). The latter is motivated by the fact that the 6th edition of the Sourcebook \parencite{aebi2021}, the last to be fully published, covers the years 2011-2016.\footnote{While a preliminary 7th edition is available, we do not consider it as its underlying data is unpublished at the time of writing.} Multiple previous editions of the Sourcebook have been published. A 5th edition \parencite{aebi2017} covers the years 2007-2011 while a 4th edition \parencite{aebi2010} covers the years 2003-2007. Note that each edition of the Sourcebook overlaps slightly with its predecessor. This can cause contradictions. For example, the 5th edition reports zero convictions in Slovenia in 2011. Meanwhile, the 6th edition reports one conviction in Slovenia in 2011. Whenever contradictions arise, we use data from the most recent edition.\footnote{For a given country and year, one edition might report a missing observation. Meanwhile, another edition might report an observation. In this case, we use the available observation, regardless of whether it comes from the most recent edition. We note that there are substantial ``jumps'' in $POL_{(c,t)}$ for some countries between different editions of the Sourcebook. This might indicate that the statistics in the Sourcebook are not fully standardized.} Each edition of the Sourcebook also contains missing observations. Thus, our panel data becomes imbalanced. Given 28 countries and a data collection period of 11 years, we have 308 possible observations. However, we only have 245 observations of $POL_{(c,t)}$ and 179 observations of $CON_{(c,t)}$. The missing conviction data is unlikely to be random and is a central limitation of our study. Missingness may reflect institutional capacity, statistical quality, legal tradition, enforcement intensity, and national recording practices. Figures \ref{fig:country} and \ref{fig:year} illustrate the number of observations of $CON_{(c,t)}$ per country and year in our dataset. As can be seen, there are some countries for which we have no observations. Still, our dataset covers Northern, Southern, Eastern, and Western Europe. As can also be seen, the number of observations increases towards the end of our data collection period.  We believe this is due to improvements in the way data was collected in the Sourcebook. We refer to Appendix \ref{appendix:missingness} for an analysis of the missing values in our dataset and some robustness checks on our main findings in this regard.\footnote{Data on the United Kingdom (UK), a member of the EU until 2020, is somewhat ambiguous in the Sourcebook as statistics are provided on ``UK: England \& Wales,'' ``UK: Northern Ireland,'' and ``UK: Scotland.'' We simply sum these when possible in a clear and consistent way (this involves summing over empty entries for ``UK: Northern Ireland'' using the 6th edition). In Section \ref{sec:results}, we run extensive robustness checks, sequentially excluding countries from our models to check if our results still hold.} In our models, tables, and graphs, we measure $POL_{(c,t)}$ and $CON_{(c,t)}$ per 100,000 capita. We note that the Sourcebook  also contains statistics on convictions for some predicate crimes to money laundering. We consider these in Appendix \ref{appendix:predicate}.  

\vfill

\section{Methodology} \label{sec:method}
We use pooled and fixed-effects regression models to investigate if STR volumes in a country can be associated with the future number of money-laundering convictions in said country. Both types of models are standard in econometric analysis (see, for example, \textcite{verbeek2017,cunningham2021causal}). We stress that the different models are used to answer different questions (see the following subsections).

Cases about money laundering can take multiple years to investigate and go through criminal justice systems. Motivated by this, we use a time lag to model the relationship between STRs and convictions for money laundering.  To be specific, we consider whether STRs are associated with convictions up to and including five years after they have been filed.\footnote{According to the \textcite{EUJusticeScoreboard2022}, court cases on money laundering in most EU countries on average took around 300 days or less in 2014. In some countries, however, the average was over 600 days. Allowing additional time for investigation, we believe five years is a reasonable time-frame to consider. We stress that we do not claim STRs are useful only within five years of being filed. Rather, we simply claim that any principal association between an increase in STRs and convictions should be visible within five years.} Recall that $STR_{(c,t)}$ denotes the number of STRs in country $c$ in year $t$ and that $CON_{(c,t)}$ denotes the number of convictions in country $c$ in year $t$. Now, let $CON_{(c,t+j)}$ denote the number of convictions in country $c$ in year $t+j$, i.e., $j$ years after $t$, and let
\begin{equation}
    CON_{(c,t+1:H)}=\frac{1}{H}\sum_{j=1}^{H}CON_{(c,t+j)}
\end{equation}
denote the average number of convictions in country $c$ per year up to and including $H$ years after $t$. We measure $CON_{(c,t+j)}$ and $CON_{(c,t+1:H)}$ per 100,000 capita using $POP_{(c,t)}$.\footnote{This is based on the idea that $CON_{(c,t+j)}$ and $CON_{(c,t+1:H)}$ are associated with  $STR_{(c,t)}$, i.e., the number of STRs in year $t$, not years $t+1,...,t+H$. For robustness, we also tried measuring $CON_{(c,t+j)}$ and $CON_{(c,t+1:H)}$ using population counts in year $t+j$ and the average population count over years $t+1,...,t+H$. This does not materially change our conclusions.} We consider and model:
\begin{enumerate}
    \item the relationship between $STR_{(c,t)}$ and $CON_{(c,t+j)}$ for $j=1,2,3,4,5$ and
    \item the relationship between $STR_{(c,t)}$ and $CON_{(c,t+1:H)}$ for $H=3$ and $H=5$.
\end{enumerate}

We stress that modeling $CON_{(c,t+1:H)}$ is functionally equivalent to modeling the accumulated number of convictions $H=3$ and $H=5$ years into the future (simply dividing the accumulated counts by 3 and 5). This is motivated by an attempt to (imperfectly) capture time-distributed STR effects (i.e., how STR amounts in one year may be associated with convictions multiple years in the future). The specification, however, introduces serial correlation between observations. The specification does not directly model repeated-report dynamics or cumulative reporting inputs (e.g., how STR volumes over multiple years may be associated with convictions in a single year).

In the following subsections, we present our pooled and fixed-effects regression models. Throughout the subsections, we use $CON_{(c,t+j)}$ as the dependent variable. However, our models for $CON_{(c,t+j)}$ and $CON_{(c,t+1:H)}$ are identical; the reader may simply substitute the dependent variable. Our models use a power-law specification inspired by the Cobb-Douglas production function. The specification is motivated by the idea that convictions are a function of (i) information (i.e., STRs; corresponding to ``capital'' in the Cobb-Douglas production function) and (ii) labor (i.e., investigative resources; ``labor'' in the Cobb-Douglas production function). Note that we, for the latter, use police officers as a (rough and highly imperfect) control variable. The power-law specification is also useful because of the highly skewed nature of STR data. However, it is not the only possible specification. Appendix \ref{appendix:no_log_models} examines simple linear models and explains why we find them less preferable. The nature of our (count-based) data also invites Poisson models; as a robustness check, Appendix \ref{appendix:poisson} considers one simple such model. 

Subsection \ref{subsec:Pooled Linear Regression} presents our pooled regression models. Subsection \ref{subsec:Fixed Effects Regression}  presents our fixed-effects regression models. Note that the number of observations used to fit each model naturally changes depending on the time period ($j$ and $H$) being considered. Throughout Section \ref{sec:results}, we always specify the number of observations used to fit each model. Our implementation relies on the \texttt{statsmodels} \parencite{statsmodels} and \texttt{linearmodels} \parencite{linearmodels} Python libraries. A replication package with our data and code can be found using  \href{https://github.com/TuffvesonJensen/STR2CON}{this link}.

\subsection{Pooled Regression} \label{subsec:Pooled Linear Regression}
 Pooled regression treats panel data as cross-sectional (i.e., as if all observations are independent; corresponding to ordinary least squares regression). While unrealistic, this is a useful starting point for our analysis. We use pooled models to investigate if countries and years with more STRs (\textit{prima facie}) can be associated with more future convictions. Our pooled regression models specify the relationship between  $STR_{(c,t)}$ and $CON_{(c,t+j)}$ as
\begin{equation}
   CON_{(c,t+j)}=A\times STR_{(c,t)}^\beta \label{eq:simple_raw}
\end{equation}
where $A\in\mathbb{R}_{>0}$ measures a ``scale of production'' and $\beta\in\mathbb{R}$ measures how convictions respond to a change in STRs.\footnote{As an alternative to the power-law formulation, we also considered a simple linear formulation; see Appendix \ref{appendix:no_log_models}.} To fit the model, we use a log-transformation, yielding an ordinary least squares regression
\begin{equation}
   \log_{10}(CON_{(c,t+j)})= \alpha + \beta\times\log_{10} (STR_{(c,t)}) \label{eq:simple_log}
\end{equation}

\noindent where $\alpha=\log_{10}(A)$ is an intercept and $\beta$ is a regression coefficient.\footnote{Our dataset has 9 observations of $CON_{(c,t)}=0$ which cannot undergo a log-transformation. When we fit our models, we dynamically drop such observations. In our data, no observations of $CON_{(c,t+1:H)}$ are equal to zero (being averages over multiple years). Thus, no observations of $CON_{(c,t+1:H)}$ are dropped. As an alternative to ``drop-handling,'' we tried adding $0.01$ to all $CON_{(c,t+j)}$ observations (after taking per capita rates but before log-transforming). This does not change our material conclusions. \label{footnote:drop}} The model pools all our observations, treating them as cross-sectional.  In particular, this means that the model ignores unobserved country-specific differences (e.g., different laws regarding when and how STRs should be filed).  This may bias our results.

Motivated by the observation that we have per capita ratios on both sides of equation (\ref{eq:simple_log}) (note that both $CON_{(c,t+j)}$ and $STR_{(c,t)}$ are measured per 100,000 capita) and that a country's population size might have a stand-alone effect on conviction rates, we include $log_{10}(POP_{(c,t)})$ as a control variable in the model in (\ref{eq:simple_log}).\footnote{Algebraically, $\log_{10}(y/n)=\alpha+\beta\,\log_{10}(x/n)+u\iff \log_{10} y=\alpha+\beta\,\log_{10}(x)+(1-\beta)\,\log_{10} (n)+u$. Thus, omitting $\log_{10}(n)$ (or, in our case, $\log_{10}(POP_{(c,t)})$) as a control variable imposes a (hidden) restriction that its coefficient must equal $1-\beta$. Note, also, that measuring variables per (100,000) capita shifts the intercept of our models.} Furthermore, we sequentially log-transform and add additional control variables for robustness. In particular, the amount of money laundering in a country could, in and of itself, drive both STRs and convictions. Imagine, for example, that banks consistently file STRs on 10\% of all money launderers and that 5\% of all launderers are convicted regardless of the number of STRs filed. In this case, an increase in launderers ``mechanically'' leads to more STRs and convictions,  creating omitted variable bias. To address this, we would like to control for the amount of money laundering in a country; something we, for obvious reasons, do not have data on. Instead, we add $\log_{10}(SGDP_{(c,t)})$ as a control variable (see the discussion in Section \ref{subsec:IMF}).\footnote{The reader may be concerned that $SGDP_{(c,t)}$ builds on an estimate from a macroeconomic model and is not directly observed. We agree with this concern; see the discussion in Section \ref{subsec:IMF}. This means that uncertainty about $SGDP_{(c,t)}$  is not properly reflected in our regression models. However, we stress that the variable only is used as control in a subset of our models.} Motivated by the idea that convictions for money laundering may also depend on the number of police officers in a country, we also add $\log_{10}(POL_{(c,t)})$ as a control. Finally,  we also try to control for time using, respectively, (i) $t$ (i.e., time) as a regressor (assuming a log-linear time trend) and (ii) time dummy variables.\footnote{Our model in (i) may, assuming no other control variables, be specified as $\log_{10}(CON_{(c,t+j)})= \alpha + \beta\times\log_{10} (STR_{(c,t)})+\delta\times t$ where $\delta\in\mathbb{R}$ is a regression coefficient. Similarly, our model in (ii) may be specified as $\log_{10}(CON_{(c,t+j)}) = \alpha + \beta \times \log_{10}(STR_{(c,t)}) + \sum_{k=1}^{K}\delta_{(k)}\mathbf{1}_{\{t = k\}}$ where $K$ denotes the last year $t$ for which we have an observation of $STR_{(c,t)}$ to model and $\delta_{(1)},\dots,\delta_{(K)}$ are regression coefficients. For numerical stability, $t$ is measured relative to 2006. \label{lab:time_models}}  We dynamically drop observations if one of our control variables is missing (see, also, Footnote \ref{footnote:drop}). Unless otherwise stated, standard errors are clustered at the country level as implemented by \texttt{statsmodels} \parencite{statsmodels}.

\subsection{Fixed-Effects Regression} 
\label{subsec:Fixed Effects Regression}
 Our country-specific fixed-effect models try to account for unobserved country-specific differences that are stable over time. These models are used to ask if, within the same country, increases in the STR volume can be associated with more future convictions. Our fixed effect models specify the relationship between  $STR_{(c,t)}$ and $CON_{(c,t+j)}$ as
\begin{equation}
   CON_{(c,t+j)}=A_{(c)}\times STR_{(c,t)}^\beta \label{eq:raw}
\end{equation}
where $A_{(c)}\in\mathbb{R}_{>0}$ is a country-specific fixed effect and $\beta\in\mathbb{R}$ measures how convictions are  associated with a change in STRs. The term $A_{(c)}$ should not be read as a direct measure of how ``good'' a country is at converting STRs into convictions. It captures stable country-level differences in the baseline relationship between STRs and convictions. These differences may include FIU organization, legal definitions of money laundering, reporting obligations, supervisory culture, feedback mechanisms, and prosecutorial priorities.  To fit the model, we use a log-transformation, yielding a fixed-effects regression
\begin{equation}
   \log_{10}(CON_{(c,t+j)})= \alpha_{(c)} + \beta\times\log_{10} (STR_{(c,t)}) \label{eq:log}
\end{equation}
where $\alpha_{(c)}=\log_{10}(A_{(c)})$ is a country-specific fixed effect and $\beta$ is a regression coefficient. For estimation, one could use country-specific dummy variables to represent $\alpha_{(c)}$. However, for numerical reasons, it is standard to demean observations at the unit level; see, e.g., \textcite{cunningham2021causal}. The model naturally controls for unobserved country-specific  differences  that are constant over time.  Unless otherwise stated, we always cluster standard errors at the unit (i.e., country) level as implemented by \texttt{linearmodels} \parencite{linearmodels}.

As in Subsection \ref{subsec:Pooled Linear Regression}, we add $\log_{10}(POP_{(c,t)})$, $\log_{10}(SGDP_{(c,t)})$, and $\log_{10}(POL_{(c,t)})$ as control variables to the model in (\ref{eq:log}). We also try to add time controls to our fixed-effects model. The idea is to control for both unobserved country-specific differences that are stable over time and time effects and shocks that are shared across countries. To this end, we try, respectively, to include (i) $t$ (i.e., time) as a regressor (assuming a log-linear time trend) and (ii) time fixed effects.\footnote{Our model in (i) may, assuming no other control variables, be specified as $\log_{10}(CON_{(c,t+j)}) = \alpha_{(c)} + \beta \times \log_{10}(STR_{(c,t)}) + \delta \times t$ where $\delta\in\mathbb{R}$ is a regression coefficient. Similarly, our model in (ii) may be specified as $\log_{10}(CON_{(c,t+j)}) = \alpha_{(c)} + \beta \times \log_{10}(STR_{(c,t)}) + \delta_{(t)}$ where $\delta_{(t)} \in \mathbb{R}$ is a time fixed effect. For numerical stability, $t$ is measured relative to 2006. For the model in (ii), standard errors are clustered both at the unit and time level as implemented by \texttt{linearmodels}.} Time effects may capture common shocks and shared developments such as EU-level reforms, FATF pressure, supervisory guidance, institutional learning, FIU capacity building, public enforcement scandals, or changing prosecutorial priorities. Removing these dynamics can improve associational specificity, but it may also remove variation that is substantively part of the STR-use mechanism. The time controlled models are used to ask if, within the same country, after removing stable country differences and time trends shared across countries, residual deviations in STR volumes can be associated with future conviction volumes. 

\section{Results} \label{sec:results}

We first consider results from our pooled regression models, treating all observations as independent. Next, we consider results from our fixed-effects regression models,  allowing more robust controls for unobserved country-specific confounders. Subsection \ref{subsec:pooled_regression_res} presents our pooled regression results. Subsection \ref{subsec:fixed_effect_regression_res} presents our fixed-effects regression results. Throughout both subsections, we test the robustness of our results by varying dependent variables, sequentially excluding countries, and including control variables and time effects. We generally round results to two decimal places and evaluate statistical significance at a $0.05$ $p$-level. The European Sourcebook of Crime and Criminal Justice Statistics also contains statistics on convictions for some predicate crimes to money laundering: we present results using these in Appendix \ref{appendix:predicate}.

\subsection{Pooled Regression Results} 
\label{subsec:pooled_regression_res}

Table \ref{tab:pooled_res} presents our pooled regression results, considering multiple model specifications, named P1 to P4, and multiple dependent variables as explained in Section \ref{sec:method}. In model P1, we only use $\log_{10}(STR_{(c,t)})$ as a regressor. In model P2, we include $\log_{10}(POP_{(c,t)})$ as a control variable. In model P3, we include $\log_{10}(SGDP_{(c,t)})$ as a control variable. Finally, in model P4, we include $\log_{10}(POL_{(c,t)})$ as a control variable. For illustration purposes, Figure \ref{fig:pooled_reg} depicts model P1. A Q-Q plot of residuals is displayed in Appendix \ref{appendix:QQ-plot}.

{All $\log_{10}(STR_{(c,t)})$ coefficients in Table \ref{tab:pooled_res} are statistically significant, positive, and smaller than 1; the largest being equal to $0.49$ with a standard error of $0.12$ and a 95\% confidence interval of $[0.26:0.72]$. This indicates a positive pooled association consistent with a sub-linear power law: countries and years with more STRs tend to have more future money-laundering convictions, but the association is less than proportional. We stress that this should not be interpreted causally.}\footnote{The reader may wonder why $\log_{10}(POL_{(c,t)})$ is insignificant using $\log_{10}(CON_{(c,t+1:3)})$ as the dependent variable when it is significant using $\log_{10}(CON_{(c,t+1)})$, $\log_{10}(CON_{(c,t+2)})$, or $\log_{10}(CON_{(c,t+3)})$ as the dependent variable. Note, however, that it takes three \textit{subsequent} years of observations to compute $\log_{10}(CON_{(c,t+1:3)})$, why it is fitted with fewer and different observations.}

For robustness, we sequentially exclude any one country from our dataset and re-run all models. Excluding either Austria, Belgium, Denmark, Estonia, Germany, Italy, the Netherlands, Portugal, or Slovenia causes at least one $\log_{10}(STR_{(c,t)})$ coefficient to become insignificant across models P3 and P4. As a further robustness check, we also consider all models with, respectively, (i) $t$ as a regressor (assuming a log-linear time trend) and (ii) year-specific dummy variables. Results are displayed in Tables \ref{tab:pooled_time_trend} and \ref{tab:pooled_time_dummies}. Including $t$ as a regressor generally causes a small drop in our $\log_{10}(STR_{(c,t)})$ coefficients though they mostly remain significant. When we include year-specific dummy variables, the $\log_{10}(STR_{(c,t)})$ coefficients, again, generally become a bit smaller but mostly remain significant.
\vfill
\begin{table}[H]
    \centering
    \renewcommand{\arraystretch}{0.95} 
    \rowcolors{2}{gray!25}{white}
    \setlength{\tabcolsep}{7.5pt} 
    \scriptsize
    \begin{tabular}{cccccccc}
\toprule
Model ID & $n$ & $R^{2}$ & $\alpha$ & $\log_{10}\left(STR_{(c,t)}\right)$ & $\log_{10}\left(POP_{(c,t)}\right)$ & $\log_{10}\left(SGDP_{(c,t)}\right)$ & $\log_{10}\left(POL_{(c,t)}\right)$ \\
\midrule\midrule
\multicolumn{8}{l}{Dependent variable: $\log_{10}\left(CON_{(c,t+1)}\right)$} \\
\midrule
P1 & 141 & 0.28 & -1.20*** (0.27) & 0.47*** (0.12) &  &  &  \\
P2 & 141 & 0.39 & -4.23*** (1.26) & 0.49*** (0.12) & 0.43* (0.20) &  &  \\
P3 & 141 & 0.48 & -8.07*** (1.94) & 0.38** (0.13) & 0.42* (0.18) & 1.11** (0.39) &  \\
P4 & 132 & 0.49 & -12.18*** (2.44) & 0.48*** (0.14) & 0.41* (0.17) & 1.38*** (0.35) & 1.19* (0.56) \\
\midrule
\multicolumn{8}{l}{Dependent variable: $\log_{10}\left(CON_{(c,t+2)}\right)$} \\
\midrule
P1 & 147 & 0.29 & -1.13*** (0.29) & 0.47*** (0.13) &  &  &  \\
P2 & 147 & 0.42 & -4.39*** (1.26) & 0.49*** (0.12) & 0.46* (0.20) &  &  \\
P3 & 147 & 0.52 & -8.43*** (1.88) & 0.38** (0.13) & 0.44* (0.18) & 1.19** (0.39) &  \\
P4 & 137 & 0.54 & -12.65*** (2.20) & 0.48*** (0.13) & 0.43** (0.16) & 1.45*** (0.32) & 1.24* (0.53) \\
\midrule
\multicolumn{8}{l}{Dependent variable: $\log_{10}\left(CON_{(c,t+3)}\right)$} \\
\midrule
P1 & 135 & 0.27 & -1.03*** (0.29) & 0.45*** (0.13) &  &  &  \\
P2 & 135 & 0.39 & -4.11** (1.30) & 0.47*** (0.12) & 0.43* (0.20) &  &  \\
P3 & 135 & 0.52 & -8.41*** (1.92) & 0.36** (0.13) & 0.40* (0.18) & 1.27** (0.42) &  \\
P4 & 128 & 0.58 & -13.18*** (2.03) & 0.49*** (0.13) & 0.38* (0.15) & 1.54*** (0.31) & 1.48** (0.49) \\
\midrule
\multicolumn{8}{l}{Dependent variable: $\log_{10}\left(CON_{(c,t+4)}\right)$} \\
\midrule
P1 & 125 & 0.23 & -0.90** (0.28) & 0.40** (0.13) &  &  &  \\
P2 & 125 & 0.34 & -3.88** (1.35) & 0.42*** (0.12) & 0.42* (0.21) &  &  \\
P3 & 125 & 0.49 & -8.49*** (1.99) & 0.31* (0.14) & 0.38* (0.19) & 1.36** (0.45) &  \\
P4 & 121 & 0.54 & -12.09*** (1.93) & 0.43*** (0.13) & 0.35* (0.16) & 1.48*** (0.34) & 1.27* (0.52) \\
\midrule
\multicolumn{8}{l}{Dependent variable: $\log_{10}\left(CON_{(c,t+5)}\right)$} \\
\midrule
P1 & 113 & 0.20 & -0.78** (0.29) & 0.36** (0.13) &  &  &  \\
P2 & 113 & 0.32 & -3.70** (1.39) & 0.38** (0.12) & 0.41 (0.21) &  &  \\
P3 & 113 & 0.49 & -8.51*** (2.04) & 0.27* (0.13) & 0.37 (0.19) & 1.42** (0.49) &  \\
P4 & 111 & 0.56 & -12.08*** (1.99) & 0.41** (0.13) & 0.32* (0.16) & 1.52*** (0.36) & 1.34* (0.59) \\
\midrule
\multicolumn{8}{l}{Dependent variable: $\log_{10}\left(CON_{(c,t+1:3)}\right)$} \\
\midrule
P1 & 122 & 0.29 & -1.19*** (0.27) & 0.48*** (0.12) &  &  &  \\
P2 & 122 & 0.44 & -4.77*** (1.32) & 0.48*** (0.12) & 0.51* (0.20) &  &  \\
P3 & 122 & 0.50 & -8.26*** (2.00) & 0.39** (0.13) & 0.50** (0.19) & 0.99** (0.37) &  \\
P4 & 116 & 0.50 & -10.58*** (2.77) & 0.45** (0.16) & 0.51** (0.18) & 1.15** (0.38) & 0.65 (0.65) \\
\midrule
\multicolumn{8}{l}{Dependent variable: $\log_{10}\left(CON_{(c,t+1:5)}\right)$} \\
\midrule
P1 & 80 & 0.40 & -1.07*** (0.20) & 0.45*** (0.08) &  &  &  \\
P2 & 80 & 0.51 & -3.71** (1.15) & 0.45*** (0.09) & 0.38* (0.17) &  &  \\
P3 & 80 & 0.62 & -7.94*** (1.77) & 0.36*** (0.10) & 0.38* (0.15) & 1.17*** (0.31) &  \\
P4 & 80 & 0.67 & -11.20*** (2.05) & 0.46*** (0.12) & 0.39** (0.13) & 1.33*** (0.24) & 0.99* (0.48) \\
\bottomrule
\end{tabular} 
    \caption{Pooled regression results. The table contains results for multiple model specifications, named P1 to P4, and multiple dependent variables as explained in Section \ref{sec:method}. Columns include model ID, the number of observations $n$ used to fit each model, goodness of fit $R^2$, and coefficient estimates (standard errors in parentheses). We use * to denote significance at the $0.05$ $p$-level; **  at the $0.01$ $p$-level; and *** at the $0.001$ $p$-level.}
    \label{tab:pooled_res}
\end{table}
\vfill
\newpage
\vfill
\begin{table}[p]
    \centering
    \renewcommand{\arraystretch}{1} 
    \rowcolors{2}{gray!25}{white}
    \setlength{\tabcolsep}{3.8pt} 
    \scriptsize
    \begin{tabular}{ccccccccc}
\toprule
Model ID & $n$ & $R^{2}$ & $\alpha$ & $\log_{10}\left(STR_{(c,t)}\right)$ & $\log_{10}\left(POP_{(c,t)}\right)$ & $\log_{10}\left(SGDP_{(c,t)}\right)$ & $\log_{10}\left(POL_{(c,t)}\right)$ & $ t $\\
\midrule\midrule
\multicolumn{9}{l}{Dependent variable: $\log_{10}\left(CON_{(c,t+1)}\right)$} \\
\midrule
PT1 & 141 & 0.37 & -1.46*** (0.23) & 0.43*** (0.12) &  &  &  & 0.08** (0.02) \\
PT2 & 141 & 0.47 & -4.23*** (1.21) & 0.45*** (0.11) & 0.40* (0.19) &  &  & 0.07** (0.02) \\
PT3 & 141 & 0.55 & -7.96*** (1.96) & 0.34** (0.13) & 0.38* (0.17) & 1.08** (0.41) &  & 0.07** (0.02) \\
PT4 & 132 & 0.56 & -11.15*** (2.35) & 0.43** (0.14) & 0.38* (0.16) & 1.26*** (0.37) & 0.97 (0.59) & 0.06* (0.03) \\
\midrule
\multicolumn{9}{l}{Dependent variable: $\log_{10}\left(CON_{(c,t+2)}\right)$} \\
\midrule
PT1 & 147 & 0.36 & -1.37*** (0.25) & 0.44*** (0.12) &  &  &  & 0.07** (0.02) \\
PT2 & 147 & 0.48 & -4.48*** (1.21) & 0.46*** (0.12) & 0.44* (0.19) &  &  & 0.06*** (0.02) \\
PT3 & 147 & 0.58 & -8.52*** (1.91) & 0.35** (0.13) & 0.42* (0.17) & 1.19** (0.41) &  & 0.06*** (0.02) \\
PT4 & 137 & 0.60 & -12.02*** (2.14) & 0.43** (0.14) & 0.42** (0.15) & 1.37*** (0.35) & 1.07 (0.58) & 0.06** (0.02) \\
\midrule
\multicolumn{9}{l}{Dependent variable: $\log_{10}\left(CON_{(c,t+3)}\right)$} \\
\midrule
PT1 & 135 & 0.32 & -1.23*** (0.26) & 0.42*** (0.13) &  &  &  & 0.06** (0.02) \\
PT2 & 135 & 0.44 & -4.22** (1.28) & 0.44*** (0.12) & 0.42* (0.20) &  &  & 0.06** (0.02) \\
PT3 & 135 & 0.56 & -8.51*** (1.95) & 0.33* (0.14) & 0.39* (0.18) & 1.27** (0.44) &  & 0.06** (0.02) \\
PT4 & 128 & 0.61 & -12.76*** (1.97) & 0.46*** (0.13) & 0.38* (0.15) & 1.48*** (0.33) & 1.35* (0.52) & 0.05* (0.02) \\
\midrule
\multicolumn{9}{l}{Dependent variable: $\log_{10}\left(CON_{(c,t+4)}\right)$} \\
\midrule
PT1 & 125 & 0.26 & -1.06*** (0.27) & 0.39** (0.13) &  &  &  & 0.06** (0.02) \\
PT2 & 125 & 0.37 & -3.95** (1.36) & 0.41*** (0.12) & 0.41 (0.21) &  &  & 0.05** (0.02) \\
PT3 & 125 & 0.52 & -8.57*** (2.03) & 0.30* (0.14) & 0.37 (0.19) & 1.36** (0.46) &  & 0.05** (0.02) \\
PT4 & 121 & 0.56 & -11.84*** (1.91) & 0.41** (0.14) & 0.35* (0.16) & 1.45*** (0.35) & 1.18* (0.55) & 0.05* (0.02) \\
\midrule
\multicolumn{9}{l}{Dependent variable: $\log_{10}\left(CON_{(c,t+5)}\right)$} \\
\midrule
PT1 & 113 & 0.21 & -0.87** (0.29) & 0.36** (0.13) &  &  &  & 0.04* (0.02) \\
PT2 & 113 & 0.33 & -3.77** (1.41) & 0.37** (0.12) & 0.41 (0.21) &  &  & 0.04* (0.02) \\
PT3 & 113 & 0.51 & -8.62*** (2.06) & 0.26 (0.13) & 0.37 (0.20) & 1.43** (0.49) &  & 0.04* (0.02) \\
PT4 & 111 & 0.57 & -12.04*** (1.98) & 0.39** (0.13) & 0.32* (0.16) & 1.52*** (0.37) & 1.28* (0.60) & 0.04* (0.02) \\
\midrule
\multicolumn{9}{l}{Dependent variable: $\log_{10}\left(CON_{(c,t+1:3)}\right)$} \\
\midrule
PT1 & 122 & 0.39 & -1.48*** (0.24) & 0.44*** (0.12) &  &  &  & 0.09** (0.03) \\
PT2 & 122 & 0.52 & -4.84*** (1.26) & 0.45*** (0.12) & 0.48* (0.19) &  &  & 0.08** (0.03) \\
PT3 & 122 & 0.58 & -8.33*** (2.00) & 0.36** (0.13) & 0.48** (0.17) & 0.99** (0.38) &  & 0.08** (0.03) \\
PT4 & 116 & 0.57 & -9.53*** (2.68) & 0.39* (0.16) & 0.48** (0.17) & 1.05** (0.40) & 0.37 (0.68) & 0.08** (0.03) \\
\midrule
\multicolumn{9}{l}{Dependent variable: $\log_{10}\left(CON_{(c,t+1:5)}\right)$} \\
\midrule
PT1 & 80 & 0.47 & -1.28*** (0.19) & 0.43*** (0.08) &  &  &  & 0.08* (0.03) \\
PT2 & 80 & 0.57 & -3.84*** (1.10) & 0.43*** (0.08) & 0.37* (0.16) &  &  & 0.08** (0.03) \\
PT3 & 80 & 0.68 & -7.91*** (1.71) & 0.34*** (0.10) & 0.37** (0.14) & 1.13*** (0.30) &  & 0.08** (0.03) \\
PT4 & 80 & 0.70 & -10.33*** (2.02) & 0.42*** (0.12) & 0.38** (0.13) & 1.26*** (0.25) & 0.74 (0.50) & 0.06* (0.03) \\
\bottomrule
\end{tabular} 
    \caption{Pooled regression results with $t$ (i.e., ``time'') as a regressor. The table contains results for multiple model specifications, named PT1 to PT4, and multiple dependent variables as explained in Section \ref{sec:method}. Columns include model ID, the number of observations $n$ used to fit each model, goodness of fit $R^2$, and coefficient estimates (standard errors in parentheses). We use * to denote significance at the $0.05$ $p$-level; **  at the $0.01$ $p$-level; and *** at the $0.001$ $p$-level.}
    \label{tab:pooled_time_trend}
\end{table}
\vfill
\newpage
\begin{sidewaystable}
    \centering
    \tiny
    \setlength{\tabcolsep}{3pt}
    \renewcommand{\arraystretch}{1}
    \rowcolors{2}{gray!25}{white}
    \begin{adjustbox}{max width=\textwidth}
        \begin{tabular}{cccccccccccccccc}
\toprule
Model ID & $n$ & $R^{2}$ & $\alpha$ & $\log_{10}\left(STR_{(c,t)}\right)$ & $\log_{10}\left(POP_{(c,t)}\right)$ & $\log_{10}\left(SGDP_{(c,t)}\right)$ & $\log_{10}\left(POL_{(c,t)}\right)$ & $t=1$ & $t=2$ & $t=3$ & $t=4$ & $t=5$ & $t=6$ & $t=7$ & $t=8$ \\
\midrule\midrule
\multicolumn{16}{l}{Dependent variable: $\log_{10}\left(CON_{(c,t+1)}\right)$} \\
\midrule
PTD1 & 141 & 0.38 & -1.40*** (0.25) & 0.43*** (0.12) &  &  &  & -0.03 (0.13) & 0.03 (0.16) & 0.07 (0.15) & 0.34* (0.17) & 0.34 (0.18) & 0.41* (0.19) & 0.46* (0.19) & 0.54** (0.17) \\
PTD2 & 141 & 0.48 & -4.16*** (1.24) & 0.44*** (0.11) & 0.40* (0.19) &  &  & -0.06 (0.15) & -0.03 (0.16) & 0.06 (0.16) & 0.28 (0.16) & 0.28 (0.17) & 0.34 (0.18) & 0.39* (0.18) & 0.48** (0.17) \\
PTD3 & 141 & 0.55 & -7.88*** (2.00) & 0.34* (0.14) & 0.39* (0.17) & 1.08* (0.42) &  & -0.09 (0.14) & -0.04 (0.16) & 0.05 (0.16) & 0.25 (0.15) & 0.24 (0.16) & 0.31 (0.17) & 0.37* (0.16) & 0.46** (0.16) \\
PTD4 & 132 & 0.56 & -11.06*** (2.41) & 0.42** (0.15) & 0.38* (0.16) & 1.26*** (0.38) & 0.97 (0.61) & -0.11 (0.14) & -0.04 (0.16) & 0.02 (0.16) & 0.20 (0.16) & 0.19 (0.17) & 0.31 (0.19) & 0.36 (0.19) & 0.41* (0.19) \\
\midrule
\multicolumn{16}{l}{Dependent variable: $\log_{10}\left(CON_{(c,t+2)}\right)$} \\
\midrule
PTD1 & 147 & 0.38 & -1.45*** (0.26) & 0.44*** (0.13) &  &  &  & 0.09 (0.11) & 0.09 (0.12) & 0.40* (0.17) & 0.39* (0.18) & 0.46* (0.18) & 0.52** (0.19) & 0.58** (0.18) & 0.49** (0.18) \\
PTD2 & 147 & 0.49 & -4.50*** (1.20) & 0.46*** (0.12) & 0.44* (0.19) &  &  & 0.04 (0.10) & 0.11 (0.12) & 0.36* (0.15) & 0.35* (0.16) & 0.41** (0.16) & 0.47** (0.17) & 0.54*** (0.16) & 0.45** (0.16) \\
PTD3 & 147 & 0.59 & -8.48*** (1.93) & 0.35* (0.14) & 0.41* (0.17) & 1.17** (0.42) &  & 0.06 (0.11) & 0.11 (0.13) & 0.33* (0.15) & 0.32* (0.16) & 0.39* (0.16) & 0.46** (0.17) & 0.54** (0.16) & 0.45** (0.16) \\
PTD4 & 137 & 0.60 & -11.92*** (2.21) & 0.43** (0.14) & 0.42** (0.16) & 1.36*** (0.36) & 1.05 (0.60) & 0.07 (0.12) & 0.08 (0.13) & 0.28 (0.16) & 0.26 (0.17) & 0.34 (0.18) & 0.46* (0.20) & 0.48* (0.20) & 0.44* (0.19) \\
\midrule
\multicolumn{16}{l}{Dependent variable: $\log_{10}\left(CON_{(c,t+3)}\right)$} \\
\midrule
PTD1 & 135 & 0.34 & -1.33*** (0.24) & 0.43*** (0.13) &  &  &  & 0.05 (0.10) & 0.31* (0.15) & 0.32* (0.15) & 0.40** (0.15) & 0.46** (0.16) & 0.53*** (0.15) & 0.42** (0.16) &  \\
PTD2 & 135 & 0.45 & -4.32*** (1.28) & 0.44*** (0.12) & 0.42* (0.20) &  &  & 0.10 (0.11) & 0.31* (0.13) & 0.33* (0.13) & 0.39** (0.13) & 0.46** (0.14) & 0.53*** (0.13) & 0.41** (0.13) &  \\
PTD3 & 135 & 0.57 & -8.48*** (1.97) & 0.34* (0.14) & 0.39* (0.18) & 1.24** (0.45) &  & 0.06 (0.11) & 0.25* (0.12) & 0.25* (0.12) & 0.32** (0.11) & 0.39** (0.13) & 0.48*** (0.12) & 0.39** (0.13) &  \\
PTD4 & 128 & 0.62 & -12.69*** (2.08) & 0.46*** (0.14) & 0.38* (0.16) & 1.48*** (0.34) & 1.34* (0.55) & 0.02 (0.10) & 0.15 (0.12) & 0.15 (0.12) & 0.23 (0.12) & 0.30* (0.13) & 0.39** (0.15) & 0.32* (0.15) &  \\
\midrule
\multicolumn{16}{l}{Dependent variable: $\log_{10}\left(CON_{(c,t+4)}\right)$} \\
\midrule
PTD1 & 125 & 0.28 & -1.22*** (0.25) & 0.39** (0.13) &  &  &  & 0.28 (0.16) & 0.25 (0.18) & 0.37* (0.18) & 0.43* (0.19) & 0.50** (0.19) & 0.40* (0.19) &  &  \\
PTD2 & 125 & 0.38 & -4.02** (1.40) & 0.41*** (0.12) & 0.40 (0.21) &  &  & 0.23 (0.14) & 0.20 (0.16) & 0.31 (0.16) & 0.37* (0.17) & 0.45** (0.17) & 0.34* (0.17) &  &  \\
PTD3 & 125 & 0.53 & -8.57*** (2.07) & 0.30* (0.14) & 0.37 (0.20) & 1.35** (0.47) &  & 0.19 (0.13) & 0.14 (0.15) & 0.24 (0.15) & 0.32* (0.16) & 0.41* (0.17) & 0.33* (0.16) &  &  \\
PTD4 & 121 & 0.57 & -11.87*** (1.98) & 0.41** (0.14) & 0.34* (0.17) & 1.46*** (0.36) & 1.18* (0.57) & 0.15 (0.14) & 0.08 (0.18) & 0.19 (0.18) & 0.26 (0.18) & 0.34 (0.20) & 0.30 (0.20) &  &  \\
\midrule
\multicolumn{16}{l}{Dependent variable: $\log_{10}\left(CON_{(c,t+5)}\right)$} \\
\midrule
PTD1 & 113 & 0.22 & -0.86** (0.29) & 0.36** (0.13) &  &  &  & -0.02 (0.10) & 0.07 (0.09) & 0.16 (0.10) & 0.22* (0.10) & 0.13 (0.09) &  &  &  \\
PTD2 & 113 & 0.34 & -3.76** (1.43) & 0.37** (0.12) & 0.41 (0.22) &  &  & -0.02 (0.09) & 0.06 (0.09) & 0.15 (0.10) & 0.22* (0.10) & 0.12 (0.09) &  &  &  \\
PTD3 & 113 & 0.52 & -8.59*** (2.09) & 0.26 (0.14) & 0.37 (0.20) & 1.43** (0.50) &  & -0.05 (0.09) & 0.02 (0.08) & 0.12 (0.10) & 0.20* (0.10) & 0.13 (0.08) &  &  &  \\
PTD4 & 111 & 0.58 & -12.01*** (2.02) & 0.39** (0.14) & 0.32* (0.16) & 1.53*** (0.37) & 1.28* (0.61) & -0.05 (0.10) & 0.01 (0.10) & 0.10 (0.10) & 0.17 (0.11) & 0.13 (0.09) &  &  &  \\
\midrule
\multicolumn{16}{l}{Dependent variable: $\log_{10}\left(CON_{(c,t+1:3)}\right)$} \\
\midrule
PTD1 & 122 & 0.40 & -1.47*** (0.25) & 0.44*** (0.12) &  &  &  & -0.03 (0.07) & 0.15 (0.10) & 0.36** (0.12) & 0.40* (0.20) & 0.44* (0.20) & 0.54** (0.18) & 0.56** (0.19) &  \\
PTD2 & 122 & 0.53 & -4.85*** (1.27) & 0.45*** (0.12) & 0.48* (0.19) &  &  & -0.03 (0.07) & 0.13 (0.09) & 0.37** (0.13) & 0.36* (0.17) & 0.40* (0.17) & 0.50** (0.15) & 0.52*** (0.15) &  \\
PTD3 & 122 & 0.59 & -8.30*** (2.04) & 0.36** (0.13) & 0.48** (0.18) & 0.98* (0.39) &  & -0.04 (0.07) & 0.13 (0.10) & 0.35* (0.15) & 0.33 (0.17) & 0.38* (0.17) & 0.49** (0.16) & 0.51** (0.16) &  \\
PTD4 & 116 & 0.58 & -9.54*** (2.74) & 0.39* (0.16) & 0.48** (0.18) & 1.05* (0.41) & 0.38 (0.68) & -0.05 (0.07) & 0.12 (0.11) & 0.34* (0.15) & 0.31 (0.18) & 0.36* (0.18) & 0.49** (0.19) & 0.50** (0.19) &  \\
\midrule
\multicolumn{16}{l}{Dependent variable: $\log_{10}\left(CON_{(c,t+1:5)}\right)$} \\
\midrule
PTD1 & 80 & 0.47 & -1.30*** (0.20) & 0.43*** (0.08) &  &  &  & 0.11* (0.05) & 0.18* (0.07) & 0.32*** (0.08) & 0.38* (0.15) & 0.41** (0.15) &  &  &  \\
PTD2 & 80 & 0.58 & -3.89*** (1.11) & 0.43*** (0.08) & 0.37* (0.17) &  &  & 0.10* (0.05) & 0.17* (0.07) & 0.34*** (0.09) & 0.36** (0.14) & 0.39** (0.13) &  &  &  \\
PTD3 & 80 & 0.69 & -7.95*** (1.72) & 0.34*** (0.10) & 0.38** (0.14) & 1.13*** (0.31) &  & 0.05 (0.06) & 0.11 (0.08) & 0.30** (0.10) & 0.31* (0.13) & 0.35** (0.12) &  &  &  \\
PTD4 & 80 & 0.71 & -10.38*** (2.07) & 0.42*** (0.12) & 0.38** (0.13) & 1.26*** (0.26) & 0.74 (0.50) & 0.04 (0.06) & 0.10 (0.08) & 0.28** (0.10) & 0.26 (0.13) & 0.30* (0.13) &  &  &  \\
\bottomrule
\end{tabular}
    \end{adjustbox} 
    \caption{Pooled regression results with year-specific dummy variables. We measure $t$ relative to 2006 (the start of our data collection period). Thus, ``$\textit{t}=1$'' denotes a dummy variable for 2007, ``$\textit{t}=2$'' denotes a dummy variable for 2008, and so on. The table contains results for multiple model specifications, named PTD1 to PTD4, and multiple dependent variables as explained in Section \ref{sec:method}. Columns include model ID, the number of observations $n$ used to fit each model, goodness of fit $R^2$, and coefficient estimates (standard errors in parentheses). We use * to denote significance at the $0.05$ $p$-level; **  at the $0.01$ $p$-level; and *** at the $0.001$ $p$-level.}
    \label{tab:pooled_time_dummies}
\end{sidewaystable}

\clearpage

\subsection{Fixed-Effects Regression Results} \label{subsec:fixed_effect_regression_res}
\begin{figure}[p]
    \centering    \includegraphics[width=0.95\linewidth]{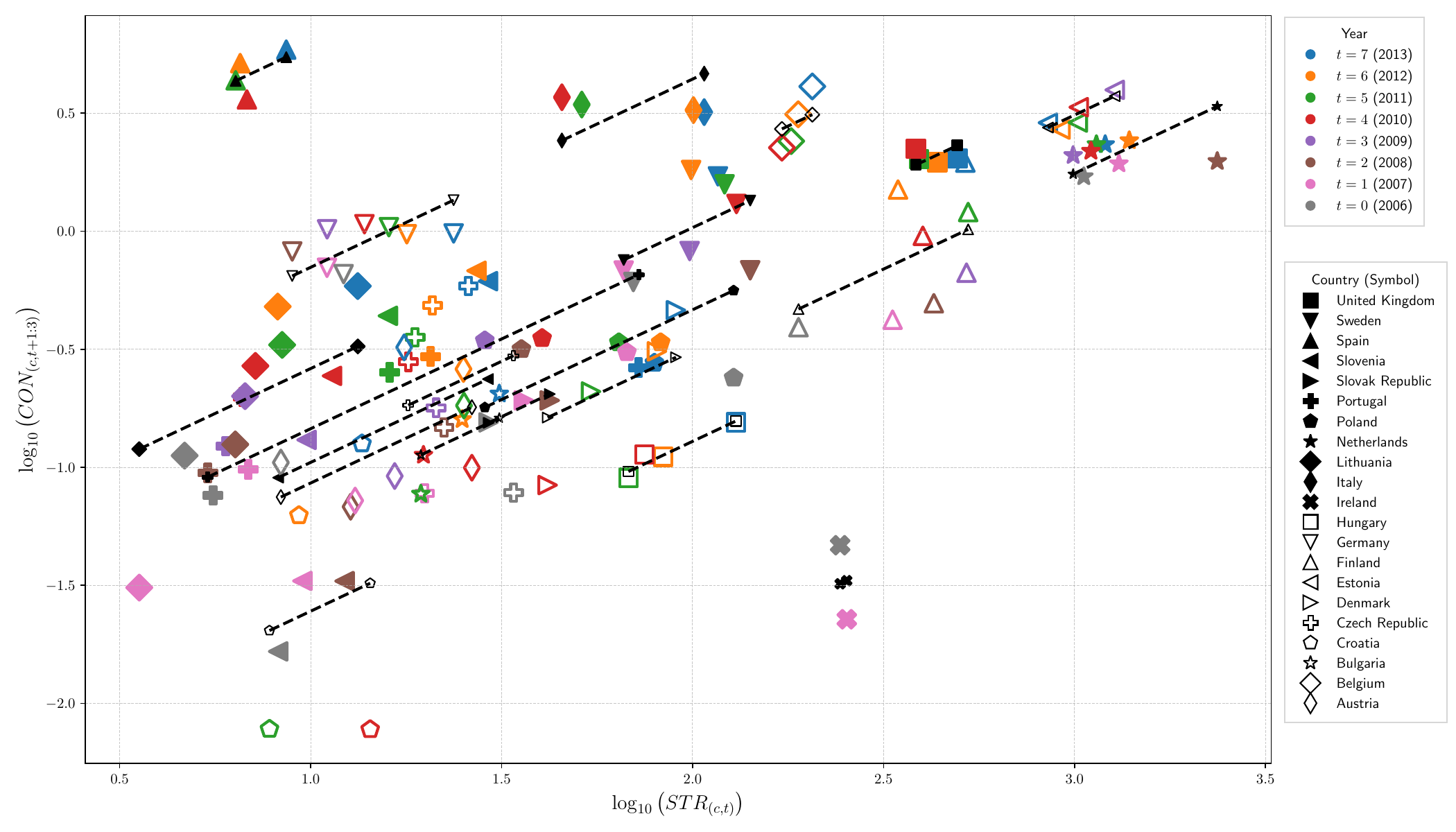} 
    \caption{Illustration of our most basic fixed-effects regression model (FE1). $STR_{(c,t)}$ denotes the number of STRs filed in country $c$ in year $t$ per 100,000 capita. $CON_{(c,t+1:3)}$ denotes the average number of money-laundering convictions per year in country $c$ up to three years after $t$ per 100,000 capita. We measure $t$ relative to 2006 (the start of our data collection). Thus, $\textit{t}=0$ corresponds to 2006, $\textit{t}=1$ corresponds to 2007, and so on.}
    \label{fig:fixed_reg}
\end{figure}

{Table \ref{tab:fixed_res} presents our fixed-effects regression results. As before, the table contains results for multiple model specifications, now named FE1 to FE4, and multiple dependent variables as explained in Section \ref{sec:method}. These models ask whether, within the same country, increases in STR volume are followed by more convictions. For illustration purposes, Figure \ref{fig:fixed_reg} depicts our most basic fixed-effects regression model (FE1) without any control variables. A Q-Q plot of residuals is displayed in Appendix A. }

All models in Table \ref{tab:fixed_res} have significant F-statistics for poolability, indicating that fixed-effects regression models are appropriate (and that pooled regression models are not). Our $\log_{10}(STR_{(c,t)})$ coefficients are statistically significant except when we use $\log_{10}(CON_{(c,t+4)})$ or $\log_{10}(CON_{(c,t+5)})$ as the dependent variable, indicating that the relationship between STRs and convictions might disappear after three years. When significant, the fixed effects $\log_{10}(STR_{(c,t)})$ coefficients are generally substantially larger than the pooled $\log_{10}(STR_{(c,t)})$ coefficients (compare Tables \ref{tab:pooled_res} and \ref{tab:fixed_res}), indicating that the relationship between STRs and convictions is affected by cross-country differences. Our largest $\log_{10}(STR_{(c,t)})$ coefficient equals $0.80$ with a standard error of $0.19$ and a $95\%$ confidence interval of $[0.42:1.18]$. While the confidence interval allows a coefficient larger than 1, our results consistently suggest a sub-linear relationship between STRs and convictions.

To complement Figure \ref{fig:fixed_reg}, Table \ref{tab:country_fe} presents estimates of the country-specific fixed effects $A_{(c)}$ and $\alpha_{(c)}$ estimated in model FE1 with $\log_{10}(CON_{(c,t+1:3)})$ as the dependent variable (i.e., the model depicted in Figure \ref{fig:fixed_reg}). Note that $\alpha_{(c)}$ is measured on a $\log_{10}$-scale (why it can be negative). The different effects could potentially reflect different legal frameworks, crime types, prosecutorial priorities, or differences in how FIUs operate. We stress that this is not investigated in our paper. The table should not be seen as a ranking; there are many reasons why countries might have different fixed effects. Values are impacted by our scaling.

For robustness, we sequentially exclude any one country from our dataset and re-run all models. Regardless of which country we exclude, the $\log_{10}(STR_{(c,t)})$ coefficients remain significant  when $\log_{10}(CON_{(c,t+1)})$, $\log_{10}(CON_{(c,t+2)})$, $\log_{10}(CON_{(c,t+3)})$, or $\log_{10}(CON_{(c,t+1:3)})$ are used as the dependent variable. Using $\log_{10}(CON_{(c,t+1:5)})$ as the dependent variable,  the $\log_{10}(STR_{(c,t)})$ coefficients become insignificant in one or more models when we exclude either Austria, Finland, Germany, Lithuania, Portugal, Slovenia, or Sweden. We also consider all models with, respectively, (i) $t$ as a regressor (assuming a log-linear time trend) and (ii) time fixed effects (alongside our country-effects). For the latter, we cluster standard errors both at the unit and time level as implemented by the \texttt{PanelOLS} class of \texttt{linearmodels} \parencite{linearmodels}. 
Table \ref{tab:fixed_time_trend} displays our results using $t$ as a regressor. All coefficients associated with $t$ are positive and significant, indicating a log-linear time trend. Meanwhile, the $\log_{10}(STR_{(c,t)})$ coefficients become much smaller and insignificant.\footnote{One might suspect that the vanishing significance of the $\log_{10}(STR_{(c,t)})$ coefficients is due to ``too little variation'' in STRs within countries. However, Figure \ref{fig:fixed_reg}, employing a log scale, shows that there is substantial variation in STRs within countries. Also, as noted in Section \ref{sec:data}, STRs per 100,000 capita within any given country on average (taken across all countries) increased by almost 240\% between 2006 and 2014 (the median increase was approximately 140\%).} Most (though not all) standard errors associated with the $\log_{10}(STR_{(c,t)})$ coefficients decrease slightly or stay approximately the same when we add $t$ as a regressor (compare Tables \ref{tab:fixed_res} and \ref{tab:fixed_time_trend}). Thus, the $\log_{10}(STR_{(c,t)})$ coefficients becoming insignificant appears to be due to reduced effect sizes, not increased standard errors. This suggests a simple possibility; an unobserved third factor, related to time, could be associated with both STRs and convictions. In Section \ref{sec:con}, we speculate that this may be related to public, political, and regulatory pressure. Note, furthermore, that all F-statistics for poolability are significant in Table \ref{tab:fixed_time_trend}. Thus, including $t$ as a regressor does not annul the significance of our country-specific fixed effects. Table \ref{tab:time_fixed_effects} displays our results when we include time-specific fixed effects in addition to our country-specific fixed effects. Again, the $\log_{10}(STR_{(c,t)})$ coefficients generally become much smaller and insignificant (or even significantly negative). This lends credibility to the possibility that an unobserved third factor, associated with time, affects both STRs and convictions.

We stress that the association between STRs and convictions only attenuates (no longer being statistically distinguishable from zero) when we control for \textit{both} country-specific and time effects. Indeed, the association is significant in Tables \ref{tab:pooled_time_trend} and \ref{tab:pooled_time_dummies}. Furthermore, the association is significant in Table \ref{tab:fixed_res}; in fact, only controlling for country-specific effects appears to increase the size of the association (compare Tables \ref{tab:pooled_res} and \ref{tab:fixed_res}). The result, however, is still that residual deviations in STR volumes do not appear to be associated with future conviction volumes once stable country differences and common time dynamics are removed. A robustness check restricted to countries with at least 10 of 11 possible conviction observations yields the same pattern; see Appendix \ref{appendix:missingness}.

\begin{table}[p]
    \centering
    \renewcommand{\arraystretch}{1} 
    \rowcolors{2}{gray!25}{white}
    \setlength{\tabcolsep}{6pt} 
    \scriptsize
    \begin{tabular}{cccccccc}
\toprule
Model ID & $n$ & $R^{2}$ (within) & F (poolability) & $\log_{10}\left(STR_{(c,t)}\right)$ & $\log_{10}\left(POP_{(c,t)}\right)$ & $\log_{10}\left(SGDP_{(c,t)}\right)$ & $\log_{10}\left(POL_{(c,t)}\right)$ \\
\midrule\midrule
\multicolumn{8}{l}{Dependent variable: $\log_{10}\left(CON_{(c,t+1)}\right)$} \\
\midrule
FE1 & 141 & 0.20 & 17.27*** & 0.71*** (0.18) &  &  &  \\
FE2 & 141 & 0.20 & 13.74*** & 0.73*** (0.17) & 3.08 (10.23) &  &  \\
FE3 & 141 & 0.21 & 11.06*** & 0.75*** (0.16) & 3.84 (9.99) & 0.82 (1.07) &  \\
FE4 & 132 & 0.25 & 10.44*** & 0.80*** (0.19) & 4.67 (11.00) & 0.83 (1.20) & -1.37 (1.24) \\
\midrule
\multicolumn{8}{l}{Dependent variable: $\log_{10}\left(CON_{(c,t+2)}\right)$} \\
\midrule
FE1 & 147 & 0.18 & 19.18*** & 0.68** (0.21) &  &  &  \\
FE2 & 147 & 0.18 & 14.56*** & 0.67*** (0.20) & -1.28 (11.09) &  &  \\
FE3 & 147 & 0.18 & 11.03*** & 0.68*** (0.19) & -0.84 (10.70) & 0.45 (1.12) &  \\
FE4 & 137 & 0.22 & 11.11*** & 0.67*** (0.19) & -0.15 (11.64) & 0.56 (1.13) & -1.74 (1.16) \\
\midrule
\multicolumn{8}{l}{Dependent variable: $\log_{10}\left(CON_{(c,t+3)}\right)$} \\
\midrule
FE1 & 135 & 0.16 & 19.67*** & 0.64** (0.21) &  &  &  \\
FE2 & 135 & 0.16 & 15.39*** & 0.64** (0.20) & 1.55 (11.65) &  &  \\
FE3 & 135 & 0.16 & 11.16*** & 0.68** (0.21) & 2.79 (10.74) & 1.27 (1.50) &  \\
FE4 & 128 & 0.17 & 9.37*** & 0.67** (0.23) & 3.35 (11.55) & 0.96 (1.65) & -0.99 (0.84) \\
\midrule
\multicolumn{8}{l}{Dependent variable: $\log_{10}\left(CON_{(c,t+4)}\right)$} \\
\midrule
FE1 & 125 & 0.02 & 17.74*** & 0.20 (0.24) &  &  &  \\
FE2 & 125 & 0.03 & 14.45*** & 0.20 (0.24) & 5.71 (12.71) &  &  \\
FE3 & 125 & 0.04 & 10.15*** & 0.22 (0.26) & 6.93 (12.17) & 1.54 (1.80) &  \\
FE4 & 121 & 0.06 & 8.91*** & 0.21 (0.28) & 6.80 (12.07) & 1.25 (1.86) & -1.31 (0.74) \\
\midrule
\multicolumn{8}{l}{Dependent variable: $\log_{10}\left(CON_{(c,t+5)}\right)$} \\
\midrule
FE1 & 113 & 0.02 & 24.66*** & 0.19 (0.26) &  &  &  \\
FE2 & 113 & 0.06 & 21.14*** & 0.18 (0.26) & 8.39 (7.06) &  &  \\
FE3 & 113 & 0.07 & 14.49*** & 0.19 (0.26) & 9.17 (7.31) & 1.39 (1.41) &  \\
FE4 & 111 & 0.09 & 12.20*** & 0.24 (0.29) & 8.40 (6.88) & 1.51 (1.60) & -1.01 (0.81) \\
\midrule
\multicolumn{8}{l}{Dependent variable: $\log_{10}\left(CON_{(c,t+1:3)}\right)$} \\
\midrule
FE1 & 122 & 0.22 & 22.22*** & 0.76** (0.26) &  &  &  \\
FE2 & 122 & 0.22 & 16.58*** & 0.77** (0.26) & 3.93 (11.31) &  &  \\
FE3 & 122 & 0.22 & 14.04*** & 0.78** (0.25) & 4.19 (11.18) & 0.32 (1.05) &  \\
FE4 & 116 & 0.26 & 14.49*** & 0.76** (0.25) & 5.40 (11.78) & 0.56 (1.10) & -1.78 (1.41) \\
\midrule
\multicolumn{8}{l}{Dependent variable: $\log_{10}\left(CON_{(c,t+1:5)}\right)$} \\
\midrule
FE1 & 80 & 0.14 & 17.23*** & 0.53* (0.25) &  &  &  \\
FE2 & 80 & 0.18 & 14.09*** & 0.53* (0.26) & 8.68 (13.63) &  &  \\
FE3 & 80 & 0.23 & 10.68*** & 0.60* (0.23) & 9.38 (12.89) & 3.05** (0.98) &  \\
FE4 & 80 & 0.30 & 10.15*** & 0.58** (0.21) & 10.42 (12.95) & 3.05*** (0.77) & -2.14 (1.72) \\
\bottomrule
\end{tabular} 
    \caption{Fixed-effects regression results. The table contains results for multiple model specifications, named FE1 to FE4, and multiple dependent variables as explained in Section \ref{sec:method}. Columns include model ID, the number of observations $n$ used to fit each model, goodness of fit $R^2$ (within), F-statistic for poolability, and coefficient estimates (standard errors in parentheses). We use * to denote significance at the $0.05$ $p$-level; ** at the $0.01$ $p$-level, and *** at the $0.001$ $p$-level.}
    \label{tab:fixed_res}
\end{table}

\begin{table}[p]
    \centering 
    \rowcolors{2}{white}{gray!25}
    \setlength{\tabcolsep}{10pt} 
    \footnotesize
    \begin{tabular}{lrr}
\toprule
Country & $\alpha_{(c)}$ & $A_{(c)}$ \\
\midrule
Spain & 0.03 & 1.06 \\
Italy & -0.88 & 0.13 \\
Germany & -0.91 & 0.12 \\
Belgium & -1.27 & 0.05 \\
Lithuania & -1.34 & 0.05 \\
Sweden & -1.51 & 0.03 \\
Portugal & -1.60 & 0.03 \\
United Kingdom & -1.68 & 0.02 \\
Czech Republic & -1.69 & 0.02 \\
Slovenia & -1.74 & 0.02 \\
Estonia & -1.79 & 0.02 \\
Austria & -1.83 & 0.01 \\
Poland & -1.85 & 0.01 \\
Slovak Republic & -1.93 & 0.01 \\
Bulgaria & -1.93 & 0.01 \\
Denmark & -2.02 & 0.01 \\
Netherlands & -2.04 & 0.01 \\
Finland & -2.06 & 0.01 \\
Croatia & -2.37 & 0.00 \\
Hungary & -2.41 & 0.00 \\
Ireland & -3.31 & 0.00 \\
\bottomrule
\end{tabular}

    \caption{Estimates of country fixed effects  $\alpha_{(c)}$ and $A_{(c)}$ using model FE1 and dependent variable $\log_{10}(CON_{(c,t+1:3)})$.}
    \label{tab:country_fe}
\end{table}

\begin{table}[p]
    \centering 
    \renewcommand{\arraystretch}{1}
    \setlength{\tabcolsep}{2.7pt} 
    \rowcolors{1}{white}{gray!25}
    \scriptsize
    \begin{tabular}{ccccccccc}
\toprule
Model ID & $n$ & $R^{2}$ (within) & F (poolability) & $\log_{10}\left(STR_{(c,t)}\right)$ & $\log_{10}\left(POP_{(c,t)}\right)$ & $\log_{10}\left(SGDP_{(c,t)}\right)$ & $\log_{10}\left(POL_{(c,t)}\right)$ & $ t $\\
\midrule\midrule
\multicolumn{9}{l}{Dependent variable: $\log_{10}\left(CON_{(c,t+1)}\right)$} \\
\midrule
FET1 & 141 & 0.42 & 21.75*** & 0.16 (0.16) &  &  &  & 0.08*** (0.02) \\
FET2 & 141 & 0.42 & 17.60*** & 0.14 (0.17) & -1.79 (4.60) &  &  & 0.08*** (0.02) \\
FET3 & 141 & 0.47 & 15.77*** & -0.13 (0.19) & -7.10 (5.68) & -4.36** (1.33) &  & 0.10*** (0.02) \\
FET4 & 132 & 0.50 & 15.21*** & -0.05 (0.18) & -5.86 (6.14) & -4.34** (1.30) & -0.98 (0.81) & 0.10*** (0.02) \\
\midrule
\multicolumn{9}{l}{Dependent variable: $\log_{10}\left(CON_{(c,t+2)}\right)$} \\
\midrule
FET1 & 147 & 0.41 & 25.51*** & 0.11 (0.18) &  &  &  & 0.07*** (0.02) \\
FET2 & 147 & 0.43 & 20.39*** & 0.05 (0.20) & -6.32 (6.07) &  &  & 0.08*** (0.02) \\
FET3 & 147 & 0.47 & 16.76*** & -0.14 (0.23) & -10.61 (6.98) & -3.55* (1.51) &  & 0.09*** (0.02) \\
FET4 & 137 & 0.48 & 16.26*** & -0.07 (0.23) & -9.44 (7.84) & -3.36* (1.49) & -1.06 (0.64) & 0.09*** (0.02) \\
\midrule
\multicolumn{9}{l}{Dependent variable: $\log_{10}\left(CON_{(c,t+3)}\right)$} \\
\midrule
FET1 & 135 & 0.38 & 26.47*** & 0.11 (0.15) &  &  &  & 0.08*** (0.02) \\
FET2 & 135 & 0.39 & 21.13*** & 0.08 (0.15) & -3.42 (6.90) &  &  & 0.08*** (0.02) \\
FET3 & 135 & 0.39 & 15.30*** & 0.04 (0.19) & -4.47 (6.96) & -0.93 (2.03) &  & 0.08*** (0.02) \\
FET4 & 128 & 0.39 & 13.04*** & 0.07 (0.20) & -3.47 (7.57) & -1.12 (2.13) & -0.85 (0.63) & 0.08*** (0.02) \\
\midrule
\multicolumn{9}{l}{Dependent variable: $\log_{10}\left(CON_{(c,t+4)}\right)$} \\
\midrule
FET1 & 125 & 0.35 & 27.71*** & -0.26 (0.19) &  &  &  & 0.09*** (0.02) \\
FET2 & 125 & 0.35 & 22.66*** & -0.27 (0.19) & -0.74 (6.94) &  &  & 0.09*** (0.02) \\
FET3 & 125 & 0.35 & 16.05*** & -0.26 (0.17) & -0.48 (6.67) & 0.29 (2.42) &  & 0.09*** (0.02) \\
FET4 & 121 & 0.38 & 14.89*** & -0.24 (0.17) & -0.75 (5.84) & 0.03 (2.48) & -1.60** (0.50) & 0.09*** (0.02) \\
\midrule
\multicolumn{9}{l}{Dependent variable: $\log_{10}\left(CON_{(c,t+5)}\right)$} \\
\midrule
FET1 & 113 & 0.27 & 33.54*** & -0.09 (0.20) &  &  &  & 0.07*** (0.02) \\
FET2 & 113 & 0.27 & 27.73*** & -0.08 (0.21) & 1.91 (4.01) &  &  & 0.07*** (0.02) \\
FET3 & 113 & 0.27 & 18.88*** & -0.08 (0.21) & 1.95 (4.51) & 0.05 (1.84) &  & 0.07*** (0.02) \\
FET4 & 111 & 0.33 & 17.17*** & -0.03 (0.21) & 0.15 (3.66) & 0.04 (2.01) & -1.62** (0.59) & 0.07*** (0.02) \\
\midrule
\multicolumn{9}{l}{Dependent variable: $\log_{10}\left(CON_{(c,t+1:3)}\right)$} \\
\midrule
FET1 & 122 & 0.52 & 33.03*** & 0.18 (0.16) &  &  &  & 0.09*** (0.02) \\
FET2 & 122 & 0.52 & 24.65*** & 0.17 (0.17) & -1.23 (5.49) &  &  & 0.09*** (0.02) \\
FET3 & 122 & 0.57 & 23.52*** & -0.04 (0.18) & -5.52 (5.66) & -4.26** (1.27) &  & 0.10*** (0.02) \\
FET4 & 116 & 0.59 & 24.37*** & -0.01 (0.18) & -3.91 (6.22) & -4.09** (1.38) & -1.37 (0.92) & 0.10*** (0.02) \\
\midrule
\multicolumn{9}{l}{Dependent variable: $\log_{10}\left(CON_{(c,t+1:5)}\right)$} \\
\midrule
FET1 & 80 & 0.63 & 37.39*** & 0.16 (0.11) &  &  &  & 0.09*** (0.02) \\
FET2 & 80 & 0.63 & 29.42*** & 0.17 (0.10) & 3.46 (6.06) &  &  & 0.09*** (0.02) \\
FET3 & 80 & 0.64 & 21.50*** & 0.10 (0.11) & 2.65 (6.25) & -1.63 (1.19) &  & 0.09*** (0.02) \\
FET4 & 80 & 0.71 & 25.04*** & 0.08 (0.08) & 3.68 (5.87) & -1.66 (1.24) & -2.24 (1.16) & 0.09*** (0.01) \\
\bottomrule
\end{tabular} 
    \caption{Fixed-effects regression results with $t$ (i.e., ``time'') included as a regressor. The table contains results for multiple model specifications, named FET1 to FET4, and multiple dependent variables. Columns include model ID, the number of observations $n$ used to fit each model, goodness of fit $R^2$ (within), F-statistic for poolability, and coefficient estimates (standard errors in parentheses). We use * to denote significance at the $0.05$ $p$-level; ** at the $0.01$ $p$-level, and *** at the $0.001$ $p$-level.}
    \label{tab:fixed_time_trend}
\end{table}

\begin{table}[p]
    \centering
    \renewcommand{\arraystretch}{1} 
    \setlength{\tabcolsep}{6pt} 
    \rowcolors{1}{white}{gray!25}
    \scriptsize
    \begin{tabular}{cccccccc}
\toprule
Model ID & $n$ & $R^{2}$ (within) & F (poolability) & $\log_{10}\left(STR_{(c,t)}\right)$ & $\log_{10}\left(POP_{(c,t)}\right)$ & $\log_{10}\left(SGDP_{(c,t)}\right)$ & $\log_{10}\left(POL_{(c,t)}\right)$ \\
\midrule\midrule
\multicolumn{8}{l}{Dependent variable: $\log_{10}\left(CON_{(c,t+1)}\right)$} \\
\midrule
FETF1 & 141 & 0.08 & 18.84*** & 0.16 (0.17) &  &  &  \\
FETF2 & 141 & 0.07 & 15.24*** & 0.14 (0.18) & -1.46 (5.49) &  &  \\
FETF3 & 141 & -0.09 & 13.88*** & -0.13 (0.17) & -6.90 (6.32) & -4.41*** (0.85) &  \\
FETF4 & 132 & -0.04 & 12.85*** & -0.05 (0.16) & -5.59 (6.58) & -4.36*** (0.73) & -1.01 (0.97) \\
\midrule
\multicolumn{8}{l}{Dependent variable: $\log_{10}\left(CON_{(c,t+2)}\right)$} \\
\midrule
FETF1 & 147 & 0.06 & 21.08*** & 0.13 (0.18) &  &  &  \\
FETF2 & 147 & 0.04 & 17.04*** & 0.08 (0.19) & -6.11 (6.11) &  &  \\
FETF3 & 147 & -0.07 & 14.53*** & -0.11 (0.21) & -10.75 (6.87) & -3.75* (1.45) &  \\
FETF4 & 137 & 0.01 & 13.83*** & -0.04 (0.23) & -9.30 (7.72) & -3.47* (1.55) & -1.26* (0.61) \\
\midrule
\multicolumn{8}{l}{Dependent variable: $\log_{10}\left(CON_{(c,t+3)}\right)$} \\
\midrule
FETF1 & 135 & 0.06 & 21.62*** & 0.13 (0.14) &  &  &  \\
FETF2 & 135 & 0.04 & 17.41*** & 0.11 (0.14) & -3.27 (6.96) &  &  \\
FETF3 & 135 & 0.01 & 13.22*** & 0.04 (0.17) & -5.22 (7.55) & -1.78 (2.15) &  \\
FETF4 & 128 & 0.03 & 11.27*** & 0.07 (0.17) & -4.02 (7.99) & -2.12 (2.29) & -1.17 (0.81) \\
\midrule
\multicolumn{8}{l}{Dependent variable: $\log_{10}\left(CON_{(c,t+4)}\right)$} \\
\midrule
FETF1 & 125 & -0.07 & 22.90*** & -0.26 (0.15) &  &  &  \\
FETF2 & 125 & -0.08 & 18.74*** & -0.26 (0.16) & -0.44 (5.93) &  &  \\
FETF3 & 125 & -0.09 & 13.67*** & -0.28* (0.11) & -1.12 (6.23) & -0.76 (2.35) &  \\
FETF4 & 121 & -0.08 & 13.12*** & -0.29* (0.13) & -1.75 (4.66) & -1.49 (2.10) & -1.89* (0.74) \\
\midrule
\multicolumn{8}{l}{Dependent variable: $\log_{10}\left(CON_{(c,t+5)}\right)$} \\
\midrule
FETF1 & 113 & -0.02 & 27.14*** & -0.08 (0.20) &  &  &  \\
FETF2 & 113 & -0.01 & 22.49*** & -0.08 (0.21) & 1.71 (5.29) &  &  \\
FETF3 & 113 & -0.00 & 15.56*** & -0.07 (0.21) & 1.95 (6.08) & 0.28 (1.89) &  \\
FETF4 & 111 & -0.00 & 14.15*** & -0.02 (0.26) & -0.36 (3.95) & -0.04 (2.12) & -1.72* (0.82) \\
\midrule
\multicolumn{8}{l}{Dependent variable: $\log_{10}\left(CON_{(c,t+1:3)}\right)$} \\
\midrule
FETF1 & 122 & 0.10 & 28.53*** & 0.20 (0.19) &  &  &  \\
FETF2 & 122 & 0.09 & 21.67*** & 0.19 (0.19) & -1.12 (5.62) &  &  \\
FETF3 & 122 & -0.03 & 21.18*** & -0.03 (0.17) & -6.06 (5.59) & -4.59** (1.50) &  \\
FETF4 & 116 & 0.03 & 21.39*** & 0.01 (0.15) & -4.47 (6.26) & -4.44** (1.62) & -1.40 (0.87) \\
\midrule
\multicolumn{8}{l}{Dependent variable: $\log_{10}\left(CON_{(c,t+1:5)}\right)$} \\
\midrule
FETF1 & 80 & 0.08 & 32.81*** & 0.18 (0.10) &  &  &  \\
FETF2 & 80 & 0.11 & 26.24*** & 0.19 (0.10) & 3.08 (5.95) &  &  \\
FETF3 & 80 & -0.03 & 20.45*** & 0.09 (0.07) & 1.63 (5.87) & -2.51 (1.90) &  \\
FETF4 & 80 & 0.04 & 22.67*** & 0.08 (0.04) & 2.80 (5.52) & -2.35 (1.77) & -2.22 (1.55) \\
\bottomrule
\end{tabular} 
    \caption{Fixed-effects regression results with both country- and time-specific effects. Note that standard errors are clustered both at the country and time level. The table contains results for multiple model specifications, named FETF1 to FETF4, and multiple dependent variables. Columns include model ID, the number of observations $n$ used to fit each model, goodness of fit $R^2$ (within), F-statistic for poolability, and coefficient estimates (standard errors in parentheses). We use * to denote significance at the $0.05$ $p$-level; ** at the $0.01$ $p$-level, and *** at the $0.001$ $p$-level. Note that negative $R^2$ (within) values can occur due to the way the statistic is computed; see the \texttt{linearmodels} documentation for details \parencite{linearmodels}.}
    \label{tab:time_fixed_effects}
\end{table}

\newpage

\section{Conclusion and Discussion} \label{sec:con}

Our paper investigates a simple but important question: are higher STR volumes associated with more future money-laundering convictions? To answer the question, we use data from Europol, the World Bank, the International Monetary Fund, and the European Sourcebook of Crime and Criminal Justice Statistics. Our analysis reveals two main findings:

\begin{enumerate}
    \item in our pooled and country-specific fixed-effects models, STR volumes and future money-laundering convictions appear to have a sub-linear power-law association, but
    \item when we add time controls to the fixed-effects specification, controlling for both country differences and common time dynamics, the association attenuates, no longer being statistically distinguishable from zero.
\end{enumerate}

The first finding is consistent with a  ``diminishing marginal association'' between STR volumes and future conviction volumes and is broadly in line with the ``crying wolf'' phenomenon as proposed by \textcite{takats2011}. The second finding implies that STR volumes do not independently predict future conviction volumes once stable country differences and common time dynamics are removed. We speculate (stressing that our results do not show) that this is related to public, political, and regulatory pressure. As noted in Section \ref{sec:data}, the average number of STRs in our dataset increased by almost 240\% between 2006 and 2014. We consider it unlikely that this increase reflects a corresponding change in underlying money-laundering activity alone. Rather, we believe it might be due to increased public, political, and regulatory pressure on financial institutions and other obligated entities. At the same time, it is possible that similar increased pressure may have influenced authorities and prosecutors across the EU to prioritize money-laundering cases (with the resources they have), leading to an increase in convictions not because more STRs have been filed, but because money-laundering cases are given a higher priority. This way, the situation could resemble omitted-variable bias. We stress that this is only one possible (and speculative) interpretation of the observed pattern and is not established or proved by our analysis.

We stress that our findings do not prove that STRs generally are ``without value.'' Our country-specific fixed effects may absorb differences in institutional mechanisms that determine how STRs become useful, including FIU design, legal definitions, supervisory culture, feedback mechanisms, and prosecutorial practice. Our time fixed effects may absorb developments in FATF pressure, FIU capacity building, and changing prosecutorial practices. In this sense, our fixed-effects model may introduce ``overcontrol'' (i.e., some of the removed variation may be part of the STRs-to-convictions mechanism itself). Furthermore, our models do not observe the full STR-to-conviction chain from reporting to FIU analysis, dissemination, investigation, and prosecution. A weak or non-existent association between STR and future conviction volumes, as found by us, can arise from attenuation at any point in the STR-to-conviction pathway.

Our study has several limitations. Our data is subject to comparability, missingness, and time limitations. Conviction data in the European Sourcebook is incomplete, with missingness unlikely to be random. Countries differ in STR reporting obligations, FIU practices, legal definitions of money laundering, and prosecution practices. While our models seek to capture time-distributed STR effects (i.e., how STR volumes in one year may be associated with conviction volumes multiple years in the future), they do not directly capture repeated-report dynamics or cumulative reporting inputs (e.g., how STR volumes over multiple years may be associated with conviction volumes in a single year). Our STR data covers the years 2006-2014 (while convictions are observed through 2016). The AML landscape has changed substantially since then, including through new EU legislation and the growth of online banking, fintech, and cryptocurrency services. Furthermore, we do not consider developments in STR quality (however defined).

The limitations mean that our findings should not be seen as a definitive assessment of contemporary AML systems. It is possible that institutional reforms, technological changes, or improved FIU practices have changed the association between STR and conviction volumes. Whether this is the case, however, remains an empirical question requiring more recent data. Our results caution against treating higher STR volumes, by themselves, as reliable indicators of future money-laundering convictions or, to the extent that convictions reflect outcomes, AML effectiveness.

\newpage

\appendix

\section{Q-Q Plots of Model Residuals} \label{appendix:QQ-plot}
\begin{figure}[H]
    \centering
    \begin{minipage}{0.48\linewidth} 
        \centering
        \includegraphics[width=\linewidth]{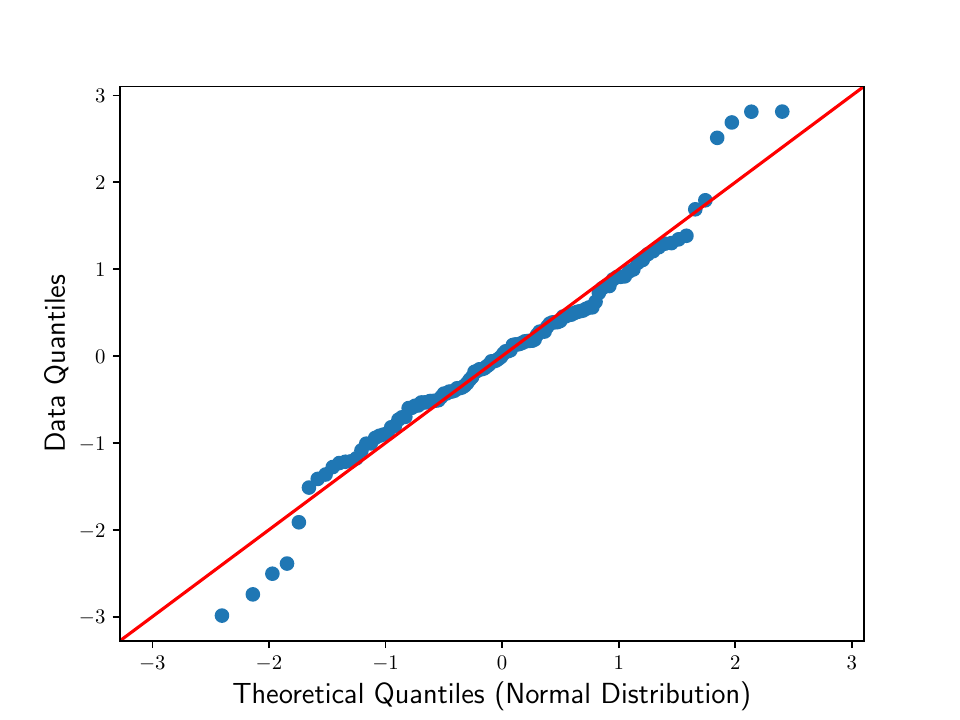}
        \caption{Q-Q plot of residuals for pooled regression model P1 using  $log_{10}(CON_{(c,t+1:3)})$  as the dependent variable.}
        \label{fig:pooled_QQ}
    \end{minipage}%
    \hfill
    \begin{minipage}{0.48\linewidth}
        \centering 
        \includegraphics[width=\linewidth]{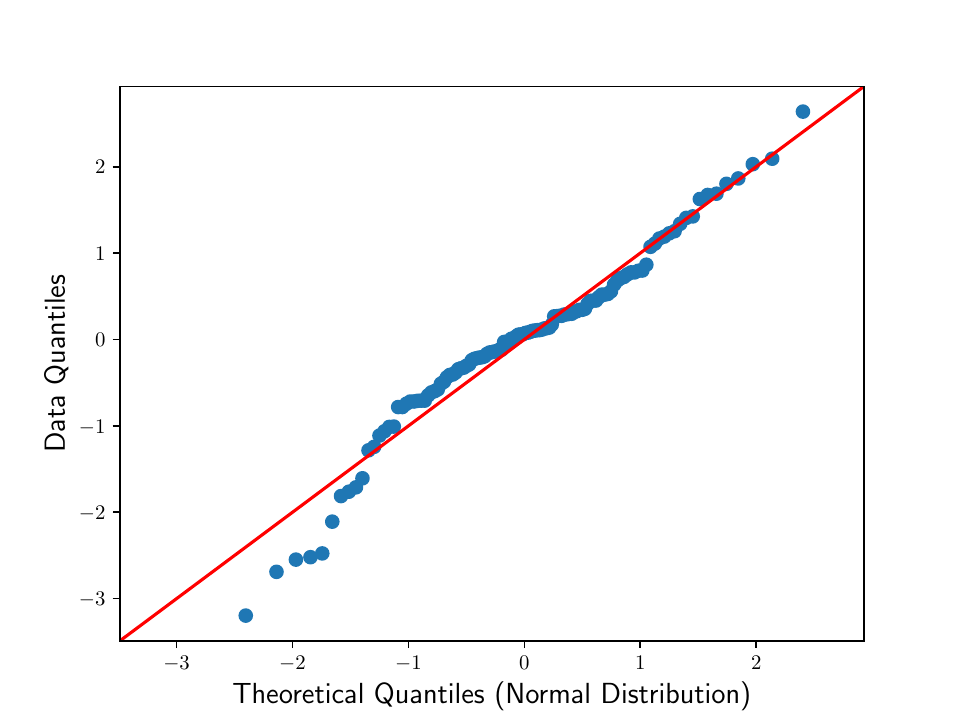}
        \caption{Q-Q plot of residuals for fixed-effects model FE1 using  $log_{10}(CON_{(c,t+1:3)})$  as the dependent variable.}
        \label{fig:fixed_QQ}
    \end{minipage}
\end{figure}

Figure \ref{fig:pooled_QQ} displays a residual Q-Q plot of our pooled regression model P1 using  $log_{10}({CON_{(c,t+1:3)}})$   as the dependent variable; see Figure \ref{fig:pooled_reg}. We note deviation from the normal distribution in both the upper and lower tails. Figure \ref{fig:fixed_QQ} displays a residual Q-Q plot of our fixed-effects regression model FE1 using $log_{10}({CON_{(c,t+1:3)}})$ as the dependent variable; see Figure \ref{fig:fixed_reg}.  We note deviation from the normal distribution in the lower tail. 

\section{Simple Linear Models without Log-transformation}
\label{appendix:no_log_models}
Inspired by the Cobb-Douglas production function, our main analysis employs a power law to model the relationship between STRs and convictions for money laundering. For completeness, we also include results using simple linear model formulations (i.e., without any log-transformation, simply regressing $CON_{(c,t+j)}$ on $STR_{(c,t)}$). Table \ref{tab:pooled_res_raw} presents results from simple linear pooled regression models, named PL1 to PL4.  Here, the $STR_{(c,t)}$ coefficients are significant (something possible even if the true relationship is a power-law). Table \ref{tab:fixed_res_raw} presents results from simple linear fixed-effects regression models, named FEL1 to FEL4. All models have significant F-statistics for poolability, indicating that fixed-effects regression models are appropriate. Meanwhile, all $STR_{(c,t)}$ coefficients are insignificant. Thus, using simple linear models, there does not appear to be a significant relationship between STRs and future convictions for money laundering when controlling for country-specific confounders.

\begin{figure}[p]
    \centering
    \includegraphics[width=0.95\linewidth]{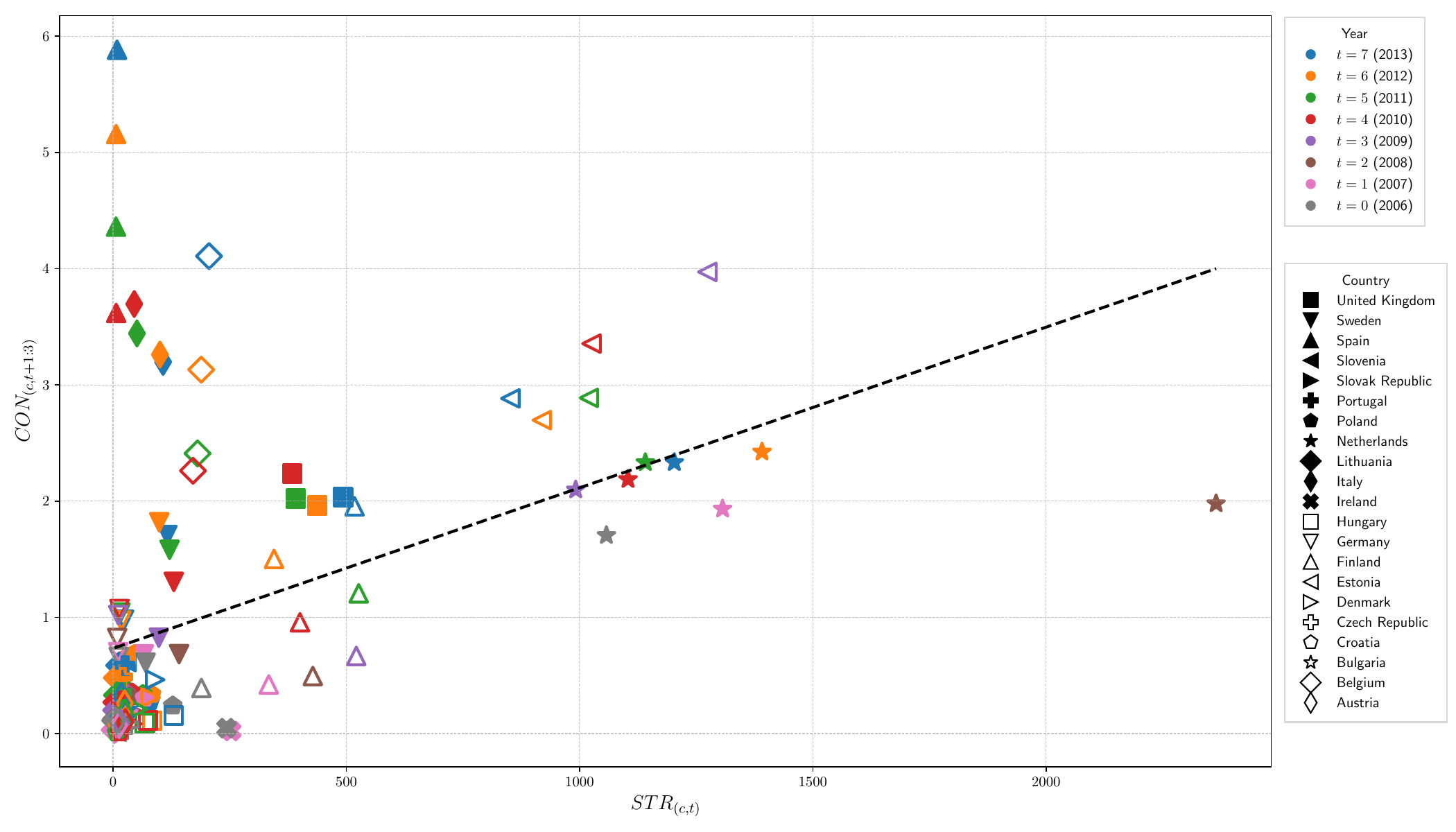} 
    \caption{Illustration of linear pooled regression model (PL1). $STR_{(c,t)}$ denotes the number of STRs filed in country $c$ in year $t$ per 100,000 capita. $CON_{(c,t+1:3)}$ denotes the average number of money-laundering convictions per year in country $c$ up to three years after $t$ per 100,000 capita. We measure $t$ relative to 2006 (the start of our data collection). Thus, $\textit{t}=0$ corresponds to 2006, $\textit{t}=1$ corresponds to 2007, and so on.}
    \label{fig:pooled_reg_raw}
\end{figure}

\begin{figure}[p]
    \centering
    \includegraphics[width=0.95\linewidth]{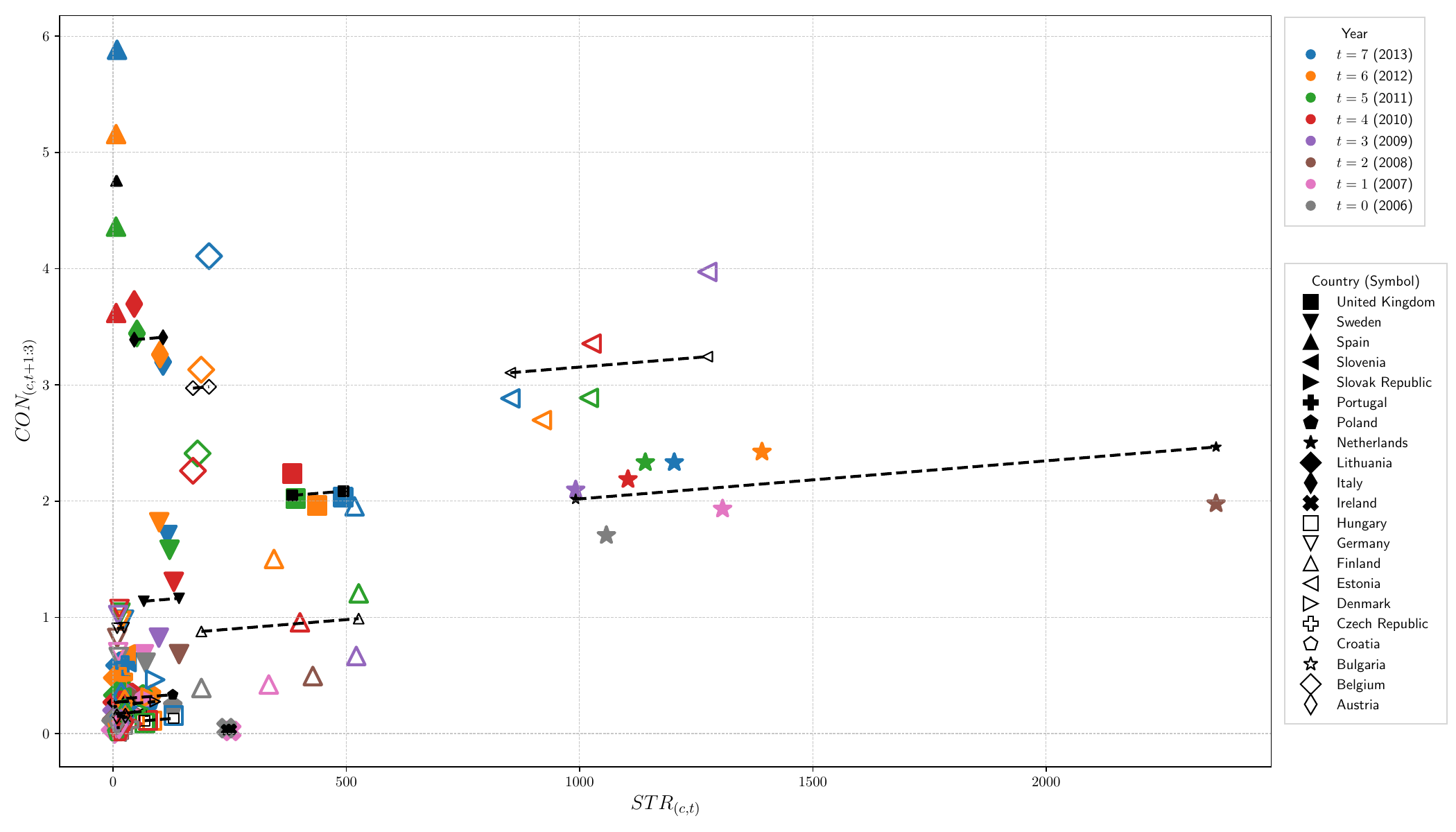} 
    \caption{Illustration of linear fixed-effects regression model (FEL1). $STR_{(c,t)}$ denotes the number of STRs filed in country $c$ in year $t$ per 100,000 capita. $CON_{(c,t+1:3)}$ denotes the average number of money-laundering convictions per year in country $c$ up to three years after $t$ per 100,000 capita. We measure $t$ relative to 2006 (the start of our data collection). Thus, $\textit{t}=0$ corresponds to 2006, $\textit{t}=1$ corresponds to 2007, and so on.}
    \label{fig:fixed_reg_raw}
\end{figure}

\begin{figure}[p]
    \centering 
    \begin{minipage}{0.48\linewidth}
        \centering
        \includegraphics[width=1\linewidth]{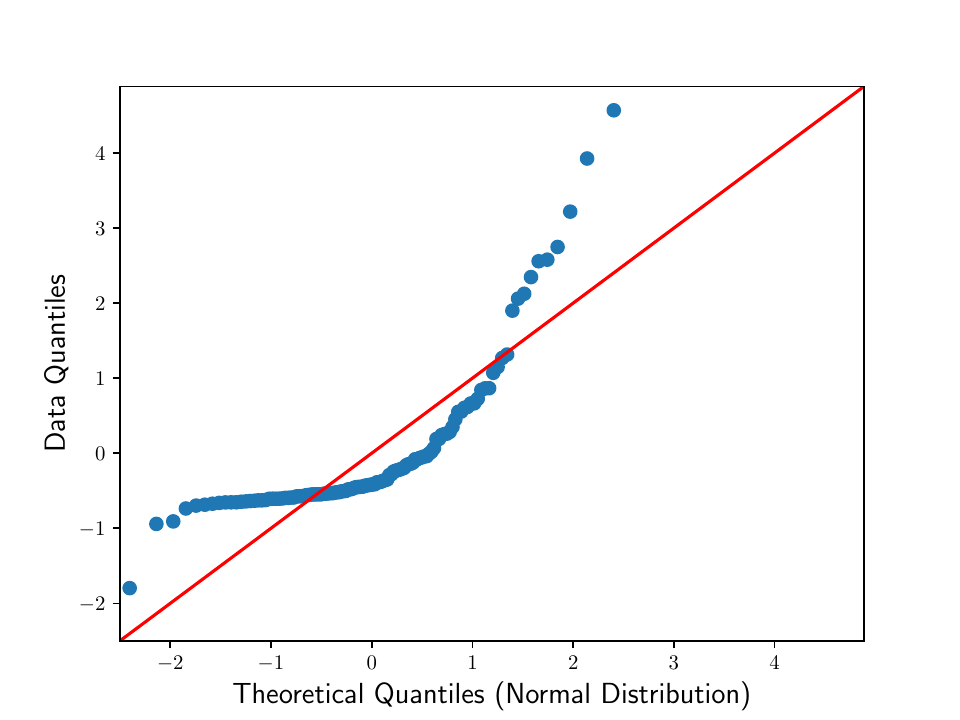}
        \caption{Q-Q plot of residuals from linear pooled regression model PL1 using $CON_{(c,t+1:3)}$ as the dependent variable.}
    \label{fig:pooled_raw_QQ}
    \end{minipage}%
    \hfill
    \begin{minipage}{0.48\linewidth}
        \centering
        \includegraphics[width=1\linewidth]{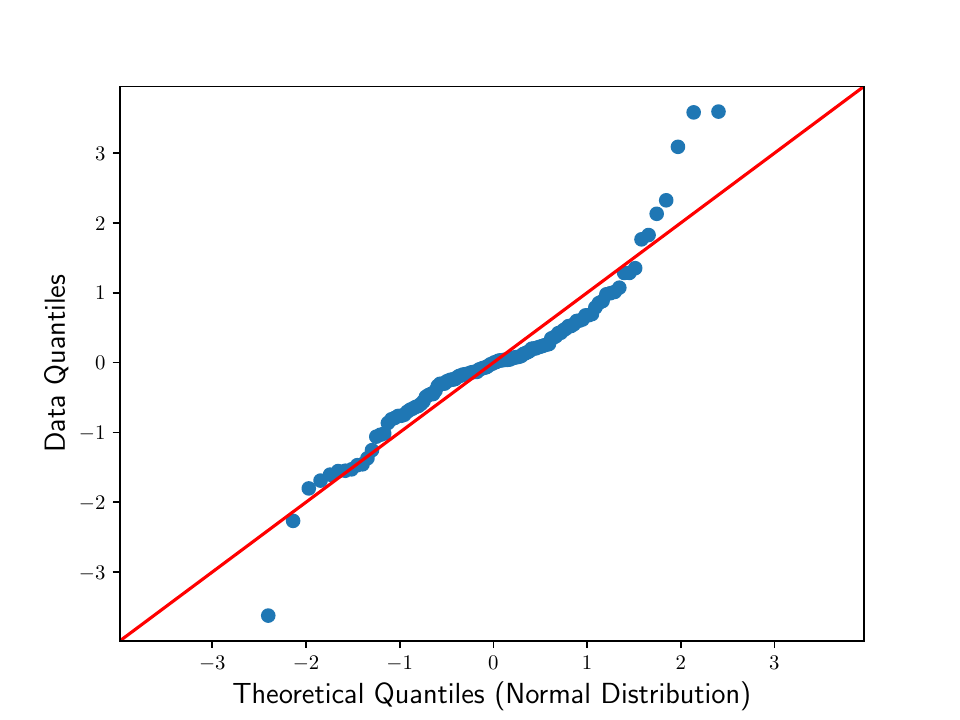}
        \caption{Q-Q plot of residuals from linear fixed-effects regression model FEL1 using $CON_{(c,t+1:3)}$ as the dependent variable.}
    \label{fig:fixed_raw_QQ}
    \end{minipage}
\end{figure}

We stress that the simple linear models are plagued by a number of problems (making us prefer the power-law formulation). Figures \ref{fig:pooled_reg_raw} and \ref{fig:fixed_reg_raw} illustrate models PL1 and FEL1 using $CON_{(c,t+1:3)}$ as the dependent variable. Here, it is easy to see how the regressions might be affected by ``outlier countries.'' Notably, we see very high condition numbers when fitting the models. Furthermore, as illustrated in Figures \ref{fig:pooled_raw_QQ} and \ref{fig:fixed_raw_QQ}, the simple linear models (especially the pooled models) suffer from (highly) non-normal residuals, showing systematic curvature in residual Q–Q plots.\footnote{This is not just a problem when we consider $CON_{(c,t+1:3)}$; it is a general problem for $CON_{(c,t+j)}$.} This makes confidence intervals and inference unreliable. Comparing, respectively, (i) Figures 
\ref{fig:pooled_reg_raw} and \ref{fig:fixed_reg_raw} with Figures \ref{fig:pooled_reg} and \ref{fig:fixed_reg} and (ii) Figures \ref{fig:pooled_raw_QQ} and \ref{fig:fixed_raw_QQ} with Figures \ref{fig:pooled_QQ} and \ref{fig:fixed_QQ}, the power-law model with a log-transformation (as used in our main analysis) is our preferred specification.
\vfill
\begin{table}[H]
    \centering 
    \rowcolors{2}{gray!25}{white}
    \tiny
    \begin{tabular}{cccccccc}
\toprule
Model ID & $n$ & $R^{2}$ & $\alpha$ & STR & POP & SGDP & POL \\
\midrule\midrule
\multicolumn{8}{l}{Dependent variable: $CON_{(c,t+1)}$} \\
\midrule
PL1 & 148 & 0.18 & 6.91e-01** (2.43e-01) & 1.41e-03** (5.22e-04) &  &  &  \\
PL2 & 148 & 0.32 & 2.82e-01 (1.65e-01) & 1.50e-03** (5.61e-04) & 2.06e-08 (1.14e-08) &  &  \\
PL3 & 148 & 0.37 & -4.10e-01 (3.51e-01) & 1.36e-03* (5.66e-04) & 1.97e-08 (1.14e-08) & 1.35e-04 (7.25e-05) &  \\
PL4 & 139 & 0.41 & -2.11e+00** (6.93e-01) & 1.63e-03** (5.72e-04) & 1.84e-08 (1.07e-08) & 1.97e-04** (6.38e-05) & 4.04e-03* (1.63e-03) \\
\midrule
\multicolumn{8}{l}{Dependent variable: $CON_{(c,t+2)}$} \\
\midrule
PL1 & 153 & 0.15 & 8.25e-01** (2.92e-01) & 1.41e-03** (5.48e-04) &  &  &  \\
PL2 & 153 & 0.29 & 3.69e-01 (2.03e-01) & 1.53e-03** (5.85e-04) & 2.22e-08 (1.26e-08) &  &  \\
PL3 & 153 & 0.36 & -5.16e-01 (3.69e-01) & 1.36e-03* (5.85e-04) & 2.08e-08 (1.25e-08) & 1.72e-04* (8.09e-05) &  \\
PL4 & 143 & 0.41 & -2.42e+00*** (6.52e-01) & 1.65e-03** (5.79e-04) & 1.92e-08 (1.16e-08) & 2.40e-04*** (6.62e-05) & 4.55e-03** (1.70e-03) \\
\midrule
\multicolumn{8}{l}{Dependent variable: $CON_{(c,t+3)}$} \\
\midrule
PL1 & 141 & 0.13 & 9.04e-01** (3.11e-01) & 1.40e-03** (5.36e-04) &  &  &  \\
PL2 & 141 & 0.28 & 4.25e-01 (2.22e-01) & 1.53e-03** (5.65e-04) & 2.30e-08 (1.31e-08) &  &  \\
PL3 & 141 & 0.36 & -5.44e-01 (3.73e-01) & 1.35e-03* (5.65e-04) & 2.13e-08 (1.29e-08) & 1.88e-04* (8.33e-05) &  \\
PL4 & 134 & 0.42 & -2.43e+00*** (5.60e-01) & 1.68e-03** (5.38e-04) & 1.90e-08 (1.13e-08) & 2.47e-04*** (6.08e-05) & 4.66e-03** (1.65e-03) \\
\midrule
\multicolumn{8}{l}{Dependent variable: $CON_{(c,t+4)}$} \\
\midrule
PL1 & 129 & 0.10 & 1.03e+00** (3.33e-01) & 1.24e-03** (4.70e-04) &  &  &  \\
PL2 & 129 & 0.24 & 5.30e-01* (2.57e-01) & 1.38e-03** (4.73e-04) & 2.36e-08 (1.36e-08) &  &  \\
PL3 & 129 & 0.35 & -5.77e-01 (3.89e-01) & 1.20e-03* (4.69e-04) & 2.14e-08 (1.32e-08) & 2.14e-04* (8.80e-05) &  \\
PL4 & 125 & 0.41 & -2.25e+00*** (6.09e-01) & 1.52e-03*** (4.21e-04) & 1.83e-08 (1.11e-08) & 2.58e-04*** (6.25e-05) & 4.33e-03* (1.83e-03) \\
\midrule
\multicolumn{8}{l}{Dependent variable: $CON_{(c,t+5)}$} \\
\midrule
PL1 & 116 & 0.08 & 1.12e+00** (3.61e-01) & 1.18e-03* (4.63e-04) &  &  &  \\
PL2 & 116 & 0.24 & 5.90e-01* (2.82e-01) & 1.32e-03** (4.46e-04) & 2.48e-08 (1.42e-08) &  &  \\
PL3 & 116 & 0.37 & -6.44e-01 (4.00e-01) & 1.14e-03* (4.44e-04) & 2.20e-08 (1.36e-08) & 2.37e-04* (9.61e-05) &  \\
PL4 & 114 & 0.45 & -2.42e+00*** (6.69e-01) & 1.53e-03*** (3.92e-04) & 1.80e-08 (1.08e-08) & 2.78e-04*** (6.81e-05) & 4.67e-03* (2.09e-03) \\
\midrule
\multicolumn{8}{l}{Dependent variable: $CON_{(c,t+1:3)}$} \\
\midrule
PL1 & 122 & 0.18 & 7.32e-01** (2.52e-01) & 1.38e-03** (4.76e-04) &  &  &  \\
PL2 & 122 & 0.32 & 3.27e-01 (1.77e-01) & 1.49e-03** (5.10e-04) & 1.99e-08 (1.15e-08) &  &  \\
PL3 & 122 & 0.38 & -4.53e-01 (3.47e-01) & 1.34e-03** (5.15e-04) & 1.90e-08 (1.15e-08) & 1.52e-04* (7.22e-05) &  \\
PL4 & 116 & 0.43 & -2.22e+00** (6.90e-01) & 1.62e-03** (5.09e-04) & 1.77e-08 (1.07e-08) & 2.17e-04*** (6.04e-05) & 4.21e-03* (1.74e-03) \\
\midrule
\multicolumn{8}{l}{Dependent variable: $CON_{(c,t+1:5)}$} \\
\midrule
PL1 & 80 & 0.23 & 6.85e-01** (2.11e-01) & 1.32e-03*** (3.99e-04) &  &  &  \\
PL2 & 80 & 0.35 & 3.45e-01* (1.64e-01) & 1.42e-03** (4.35e-04) & 1.66e-08 (9.67e-09) &  &  \\
PL3 & 80 & 0.43 & -5.02e-01 (3.26e-01) & 1.28e-03** (4.40e-04) & 1.58e-08 (9.85e-09) & 1.64e-04* (6.80e-05) &  \\
PL4 & 80 & 0.53 & -2.24e+00*** (6.64e-01) & 1.57e-03*** (4.34e-04) & 1.52e-08 (8.73e-09) & 2.28e-04*** (5.24e-05) & 4.18e-03* (1.73e-03) \\
\bottomrule
\end{tabular}
     \tiny
    \caption{Pooled regression results using a simple linear model. Columns include model ID, the number of observations $n$ used to fit each model, goodness of fit $R^2$, and coefficient estimates (standard errors in parentheses). We use * to denote significance at the $0.05$ $p$-level; ** at the $0.01$ $p$-level; 
    and *** at the $0.001$ $p$-level.}
    \label{tab:pooled_res_raw}
\end{table}
\vfill

\begin{table}[p]
    \centering 
    \rowcolors{2}{gray!25}{white}
    \tiny
    \begin{tabular}{cccccccc}
\toprule
Model ID & $n$ & $R^{2}$ (within) & F (poolability) & STR & POP & SGDP & POL \\
\midrule\midrule
\multicolumn{8}{l}{Dependent variable: $CON_{(c,t+1)}$} \\
\midrule
FEL1 & 148 & 0.09 & 36.09*** & 1.08e-03 (9.11e-04) &  &  &  \\
FEL2 & 148 & 0.10 & 28.70*** & 1.09e-03 (9.23e-04) & -9.54e-08 (1.54e-07) &  &  \\
FEL3 & 148 & 0.11 & 26.26*** & 1.06e-03 (8.72e-04) & -1.19e-07 (1.75e-07) & -2.21e-04 (3.80e-04) &  \\
FEL4 & 139 & 0.12 & 21.13*** & 1.07e-03 (8.81e-04) & -2.34e-08 (2.02e-07) & -3.53e-04 (4.07e-04) & -3.15e-04 (3.55e-03) \\
\midrule
\multicolumn{8}{l}{Dependent variable: $CON_{(c,t+2)}$} \\
\midrule
FEL1 & 153 & 0.01 & 41.07*** & 5.07e-04 (5.66e-04) &  &  &  \\
FEL2 & 153 & 0.01 & 33.02*** & 4.98e-04 (5.66e-04) & 3.92e-08 (1.43e-07) &  &  \\
FEL3 & 153 & 0.03 & 29.53*** & 4.81e-04 (5.12e-04) & -5.11e-09 (1.47e-07) & -3.09e-04 (4.20e-04) &  \\
FEL4 & 143 & 0.11 & 27.17*** & 4.95e-04 (5.44e-04) & 5.93e-08 (2.05e-07) & -3.86e-04 (3.62e-04) & -4.46e-03 (3.78e-03) \\
\midrule
\multicolumn{8}{l}{Dependent variable: $CON_{(c,t+3)}$} \\
\midrule
FEL1 & 141 & 0.00 & 37.65*** & 1.14e-04 (2.80e-04) &  &  &  \\
FEL2 & 141 & 0.02 & 30.87*** & 9.40e-05 (2.61e-04) & 2.14e-07 (2.47e-07) &  &  \\
FEL3 & 141 & 0.03 & 26.90*** & 5.73e-05 (2.28e-04) & 1.67e-07 (2.13e-07) & -2.74e-04 (4.89e-04) &  \\
FEL4 & 134 & 0.12 & 25.06*** & 1.52e-04 (2.72e-04) & 1.70e-07 (2.46e-07) & -3.38e-04 (4.57e-04) & -5.25e-03 (3.45e-03) \\
\midrule
\multicolumn{8}{l}{Dependent variable: $CON_{(c,t+4)}$} \\
\midrule
FEL1 & 129 & 0.03 & 37.90*** & -5.81e-04 (5.71e-04) &  &  &  \\
FEL2 & 129 & 0.09 & 33.12*** & -6.03e-04 (5.77e-04) & 3.87e-07 (3.49e-07) &  &  \\
FEL3 & 129 & 0.09 & 27.67*** & -6.09e-04 (5.69e-04) & 3.80e-07 (3.30e-07) & -4.14e-05 (3.53e-04) &  \\
FEL4 & 125 & 0.17 & 26.64*** & -3.91e-04 (3.10e-04) & 3.25e-07 (3.11e-07) & -7.47e-05 (3.39e-04) & -5.67e-03** (2.13e-03) \\
\midrule
\multicolumn{8}{l}{Dependent variable: $CON_{(c,t+5)}$} \\
\midrule
FEL1 & 116 & 0.01 & 37.18*** & -3.42e-04 (5.42e-04) &  &  &  \\
FEL2 & 116 & 0.12 & 33.90*** & -3.58e-04 (5.46e-04) & 4.62e-07 (3.76e-07) &  &  \\
FEL3 & 116 & 0.12 & 27.37*** & -3.28e-04 (5.08e-04) & 4.84e-07 (3.71e-07) & 2.38e-04 (2.19e-04) &  \\
FEL4 & 114 & 0.18 & 25.23*** & -1.23e-04 (2.72e-04) & 4.42e-07 (3.79e-07) & 1.32e-04 (2.67e-04) & -6.05e-03** (2.06e-03) \\
\midrule
\multicolumn{8}{l}{Dependent variable: $CON_{(c,t+1:3)}$} \\
\midrule
FEL1 & 122 & 0.01 & 59.33*** & 3.27e-04 (3.95e-04) &  &  &  \\
FEL2 & 122 & 0.02 & 48.20*** & 3.26e-04 (3.93e-04) & 5.85e-08 (1.66e-07) &  &  \\
FEL3 & 122 & 0.04 & 43.88*** & 3.29e-04 (3.54e-04) & 2.91e-08 (1.74e-07) & -2.80e-04 (3.64e-04) &  \\
FEL4 & 116 & 0.09 & 36.95*** & 3.20e-04 (3.51e-04) & 7.21e-08 (2.16e-07) & -3.62e-04 (3.63e-04) & -2.25e-03 (2.75e-03) \\
\midrule
\multicolumn{8}{l}{Dependent variable: $CON_{(c,t+1:5)}$} \\
\midrule
FEL1 & 80 & 0.02 & 83.59*** & 1.81e-04 (2.44e-04) &  &  &  \\
FEL2 & 80 & 0.03 & 70.36*** & 1.85e-04 (2.41e-04) & 1.04e-07 (1.90e-07) &  &  \\
FEL3 & 80 & 0.03 & 59.65*** & 1.85e-04 (2.60e-04) & 1.04e-07 (1.91e-07) & -3.16e-06 (1.84e-04) &  \\
FEL4 & 80 & 0.05 & 48.68*** & 1.63e-04 (2.56e-04) & 9.95e-08 (1.98e-07) & 2.46e-06 (1.92e-04) & -1.68e-03 (2.22e-03) \\
\bottomrule
\end{tabular}
    
    \caption{Fixed-effects regression results using a simple linear model. Columns include model ID, the number of observations $n$ used to fit each model, within $R^2$, and coefficient estimates (standard errors in parentheses). 
    We use * to denote significance at the $0.05$ $p$-level; ** at the $0.01$ $p$-level; 
    and *** at the $0.001$ $p$-level.}
    \label{tab:fixed_res_raw}
\end{table}

\newpage

\section{Data Missingness} \label{appendix:missingness}

Table \ref{tab:missingness_variable_level} reports variable-level missingness in our dataset. Missingness is concentrated in the conviction outcome variable $CON_{(c,t)}$ and, to a lesser extent, the police-force control variable $POL_{(c,t)}$.

\begin{table}[H]
    \centering
    \small
    \begin{tabular}{lrrrrr}
\toprule
Variable & Years & Potential obs. & Observed obs. & Missing obs. & Coverage \\
\midrule
$STR_{(c,t)}$ & 2006--2014 & 252 & 252 & 0 & 100.0\% \\
$GDP_{(c,t)}$ & 2006--2016 & 308 & 308 & 0 & 100.0\% \\
$POP_{(c,t)}$ & 2006--2016 & 308 & 308 & 0 & 100.0\% \\
$SHW_{(c,t)}$ & 2006--2016 & 308 & 297 & 11 & 96.4\% \\
$POL_{(c,t)}$ & 2006--2016 & 308 & 245 & 63 & 79.5\% \\
$CON_{(c,t)}$ & 2006--2016 & 308 & 179 & 129 & 58.1\% \\
\bottomrule
\end{tabular}

    \caption{Variable-level missingness in our dataset.}
    \label{tab:missingness_variable_level}
\end{table} 

\noindent Table \ref{tab:missingness_country_con_coverage} reports country-level coverage for our money-laundering conviction observations. Five countries have no observed convictions in our dataset: France, Greece, Latvia, Luxembourg, and Malta. Table \ref{tab:missingness_region_con_coverage} reports conviction-data coverage by region.\footnote{Countries are grouped as follows: Western Europe includes Austria, Belgium, France, Germany, Ireland, Luxembourg, the Netherlands, and the United Kingdom; Northern Europe includes Denmark, Estonia, Finland, Latvia, Lithuania, and Sweden; Eastern Europe includes Bulgaria, Croatia, the Czech Republic, Hungary, Poland, Romania, and the Slovak Republic; Southern Europe includes Cyprus, Greece, Italy, Malta, Portugal, Slovenia, and Spain.}

\begin{table}[H]
    \centering
    \scriptsize
    \begin{tabular}{lrrrr}
\toprule
Country & Observed CON years & Potential years & Missing years & Coverage \\
\midrule
Austria & 11 & 11 & 0 & 100.0\% \\
Czech Republic & 11 & 11 & 0 & 100.0\% \\
Finland & 11 & 11 & 0 & 100.0\% \\
Germany & 11 & 11 & 0 & 100.0\% \\
Lithuania & 11 & 11 & 0 & 100.0\% \\
Netherlands & 11 & 11 & 0 & 100.0\% \\
Poland & 11 & 11 & 0 & 100.0\% \\
Slovenia & 11 & 11 & 0 & 100.0\% \\
Portugal & 10 & 11 & 1 & 90.9\% \\
Sweden & 10 & 11 & 1 & 90.9\% \\
Estonia & 8 & 11 & 3 & 72.7\% \\
Belgium & 6 & 11 & 5 & 54.5\% \\
Bulgaria & 6 & 11 & 5 & 54.5\% \\
Croatia & 6 & 11 & 5 & 54.5\% \\
Denmark & 6 & 11 & 5 & 54.5\% \\
Hungary & 6 & 11 & 5 & 54.5\% \\
Italy & 6 & 11 & 5 & 54.5\% \\
Slovak Republic & 6 & 11 & 5 & 54.5\% \\
Spain & 6 & 11 & 5 & 54.5\% \\
United Kingdom & 6 & 11 & 5 & 54.5\% \\
Ireland & 4 & 11 & 7 & 36.4\% \\
Romania & 3 & 11 & 8 & 27.3\% \\
Cyprus & 2 & 11 & 9 & 18.2\% \\
France & 0 & 11 & 11 & 0.0\% \\
Greece & 0 & 11 & 11 & 0.0\% \\
Latvia & 0 & 11 & 11 & 0.0\% \\
Luxembourg & 0 & 11 & 11 & 0.0\% \\
Malta & 0 & 11 & 11 & 0.0\% \\
\bottomrule
\end{tabular}

    \caption{Country-level coverage for money-laundering conviction observations, 2006--2016.}
    \label{tab:missingness_country_con_coverage}
\end{table}

\begin{table}[H]
    \centering
    \small
    \begin{tabular}{lrrrr}
\toprule
Region & Observed CON obs. & Potential obs. & Missing obs. & Coverage \\
\midrule
Eastern Europe & 49 & 77 & 28 & 63.6\% \\
Western Europe & 49 & 88 & 39 & 55.7\% \\
Northern Europe & 46 & 66 & 20 & 69.7\% \\
Southern Europe & 35 & 77 & 42 & 45.5\% \\
\bottomrule
\end{tabular}

    \caption{Regional coverage for money-laundering conviction observations, 2006--2016.}
    \label{tab:missingness_region_con_coverage}
\end{table}

As a robustness check, Table \ref{tab:high_coverage_countries} reruns our fixed-effects model specifications for $CON_{(c,t+1:3)}$ using only countries with at least 10 of 11 possible conviction observations. This removes countries with sparse conviction coverage and tests if our main fixed-effects result is driven by countries with limited outcome data. The restricted-sample results are consistent with our main findings: the STR coefficient is positive in the country-specific fixed-effects models but is not robust to the inclusion of common time dynamics.

\begin{table}[H]
    \centering
    \tiny
    \begin{tabular}{ccccccccc}
\toprule
Model ID & $n$ & $R^{2}$ (within) & F (poolability) & $\log_{10}\left(STR_{(c,t)}\right)$ & $\log_{10}\left(POP_{(c,t)}\right)$ & $\log_{10}\left(SGDP_{(c,t)}\right)$ & $\log_{10}\left(POL_{(c,t)}\right)$ & $ t $\\
\midrule\midrule
\multicolumn{9}{l}{Dependent variable: $\log_{10}\left(CON_{(c,t+1:3)}\right)$} \\
\midrule
HCFE1 & 80 & 0.23 & 8.14*** & 0.75* (0.29) &  &  &  &  \\
HCFE2 & 80 & 0.26 & 6.08*** & 0.78* (0.29) & 6.74 (13.86) &  &  &  \\
HCFE3 & 80 & 0.26 & 3.55** & 0.80** (0.28) & 7.55 (13.37) & 1.01 (0.90) &  &  \\
HCFE4 & 78 & 0.33 & 5.13*** & 0.71** (0.24) & 8.82 (13.51) & 1.22 (0.89) & -3.04 (2.70) &  \\
HCFET1 & 80 & 0.60 & 15.25*** & 0.15 (0.16) &  &  &  & 0.09*** (0.02) \\
HCFET2 & 80 & 0.60 & 10.46*** & 0.16 (0.17) & 0.64 (6.70) &  &  & 0.09*** (0.02) \\
HCFET3 & 80 & 0.65 & 7.28*** & -0.09 (0.19) & -4.33 (6.57) & -4.63*** (1.22) &  & 0.11*** (0.02) \\
HCFET4 & 78 & 0.73 & 11.33*** & -0.16 (0.13) & -1.91 (6.46) & -4.34** (1.43) & -3.17** (1.16) & 0.11*** (0.02) \\
HCFETF1 & 80 & 0.10 & 12.28*** & 0.18 (0.20) &  &  &  &  \\
HCFETF2 & 80 & 0.11 & 9.71*** & 0.19 (0.20) & 0.77 (6.81) &  &  &  \\
HCFETF3 & 80 & -0.08 & 8.85*** & -0.05 (0.17) & -4.55 (6.41) & -4.83** (1.47) &  &  \\
HCFETF4 & 78 & 0.02 & 13.35*** & -0.11 (0.11) & -1.81 (6.49) & -4.29* (1.72) & -3.31* (1.34) &  \\
\bottomrule
\end{tabular}
    \caption{High-coverage country robustness check. The dependent variable is $\log_{10}\left(CON_{(c,t+1:3)}\right)$. High-coverage countries are defined as countries with at least 10 of 11 possible money-laundering conviction observations. Columns include model ID, the number of observations $n$ used to fit each model, goodness of fit $R^2$ (within), F-statistic for poolability, and coefficient estimates (standard errors in parentheses). Note that negative $R^2$ (within) values can occur due to the way the statistic is computed; see the \texttt{linearmodels} documentation for details \parencite{linearmodels}. We use a model ID naming convention as in our main analysis (prefixing ``HC'' for ``high-coverage''). We use * to denote significance at the $0.05$ $p$-level; ** at the $0.01$ $p$-level, and *** at the $0.001$ $p$-level.}
    \label{tab:high_coverage_countries}
\end{table}

\newpage

\section{Predicate Crimes to Money Laundering} \label{appendix:predicate} 
 Our main analysis considers the association between the number of STRs and convictions for money laundering in a country.  However, the European Sourcebook of Crime and Criminal Justice Statistics also contains statistics on some predicate crimes to money laundering, including corruption, drug trafficking and fraud. Motivated by this, we collect and model the following variables from the Sourcebook:

\begin{itemize}
    \item $CRP_{(c,t)}$: the number of people convicted for corruption (Sourcebook indicator code \texttt{T31CO}) in country $c$ in year $t$, 
    \item $DRT_{(c,t)}$: the number of people convicted for drug trafficking (Sourcebook indicator code \texttt{T31DT}) in country $c$ in year $t$, and 
    \item $FRD_{(c,t)}$: the number of people convicted for fraud (Sourcebook indicator code \texttt{T31FR}) in country $c$ in year $t$.
\end{itemize} 

\noindent We collect and measure the variables analogously to $CON_{(c,t)}$ as described in Section \ref{sec:data}. For brevity, we primarily consider fixed-effects regression models (our most robust models) in this appendix. Thus, we log-transform and model each of the variables analogously to $log_{10}(CON_{(c,t+j)})$ as described in Subsection \ref{subsec:Fixed Effects Regression}.

Table \ref{tab:fixed_res_CRP} displays results for convictions for corruption. Here, no $\log_{10}(STR_{(c,t)})$ coefficients are significant. If we consider simple linear models between $STR_{(c,t)}$ and $CRP_{(c,t+j)}$ (analogously to Appendix \ref{appendix:no_log_models}; omitted for brevity), no $STR_{(c,t)}$ coefficients are significant. Table \ref{tab:fixed_res_DRT} displays results for convictions for drug trafficking.\footnote{We do not consider drug trafficking convictions from the UK due to inconsistency in how they are denoted and available in different Sourcebook editions between, respectively, ``UK: England and Wales,'' ``UK: Northern Ireland,'' and ``UK: Scotland.''} Here, some $\log_{10}(STR_{(c,t)})$ coefficients are significant for $log_{10}(DRT_{(c,t+1)})$, $log_{10}(DRT_{(c,t+4)})$, and $log_{10}(DRT_{(c,t+5)})$. Some readers might see this as evidence that STRs affect convictions for drug-trafficking. We, however, call for caution: we are considering several outcomes (i.e., convictions for corruption, drug trafficking, and fraud) and multiple time horizons without adjusting for multiplicity or time effects. If we consider simple linear models between $STR_{(c,t)}$ and $DRT_{(c,t+j)}$ (analogously to Appendix \ref{appendix:no_log_models}; omitted for brevity), we have some significant $STR_{(c,t)}$ coefficients when considering $DRT_{(c,t+1)},DRT_{(c,t+3)}, DRT_{(c,t+4)},$ and $DRT_{(c,t+5)}$. Finally, Table \ref{tab:fixed_res_FRD} contains results for convictions for fraud.\footnote{We only consider fraud convictions from the UK when they are available and can be summed over all of ``UK: England and Wales,'' ``UK: Northern Ireland,'' and ``UK: Scotland'' in the Sourcebook.} Here, no $\log_{10}(STR_{(c,t)})$ coefficients are significant. If we consider simple linear models between $STR_{(c,t)}$ and $FRD_{(c,t+j)}$ (analogously to Appendix \ref{appendix:no_log_models}; omitted for brevity), a single model specification has a significant $STR_{(c,t)}$ coefficient when considering $FRD_{(c,t+1:5)}$. We, again, caution against over-interpretation. 

We stress that our analysis of the relationship between the number of STRs and convictions for predicate crimes is limited (especially compared to our main analysis, considering convictions for money laundering).\footnote{In particular, a robust analysis of the relationship between STRs and convictions for predicate crimes would consider time effects (as in our main analysis). This is outside of the scope of the present appendix (possibly warranting a study of its own).} We do not rule out that there is a positive relationship between STRs and convictions for predicate crimes. However, we also fail to find robust and consistent evidence of such relationships when considering convictions for corruption, drug trafficking, and fraud. The most plausible relationship might (note our cautious wording) be between the number of STRs and convictions for drug trafficking.
\vfill
\begin{table}[H]
        \centering
        \rowcolors{2}{gray!25}{white}
        \setlength{\tabcolsep}{6pt} 
        \tiny
        \begin{tabular}{cccccccc}
\toprule
Model ID & $n$ & $R^{2}$ (within) & F (poolability) & $\log_{10}\left(STR_{(c,t)}\right)$ & $\log_{10}\left(POP_{(c,t)}\right)$ & $\log_{10}\left(SGDP_{(c,t)}\right)$ & $\log_{10}\left(POL_{(c,t)}\right)$ \\
\midrule\midrule
\multicolumn{8}{l}{Dependent variable: $\log_{10}\left(CRP_{(c,t+1)}\right)$} \\
\midrule
FECRP1 & 162 & 0.03 & 85.66*** & -0.15 (0.13) &  &  &  \\
FECRP2 & 162 & 0.09 & 86.69*** & -0.17 (0.13) & -6.67 (3.96) &  &  \\
FECRP3 & 162 & 0.10 & 45.64*** & -0.17 (0.13) & -7.96 (4.03) & -1.20 (0.85) &  \\
FECRP4 & 147 & 0.16 & 43.36*** & -0.20 (0.13) & -11.13** (3.37) & -1.36 (0.91) & -0.03 (0.59) \\
\midrule
\multicolumn{8}{l}{Dependent variable: $\log_{10}\left(CRP_{(c,t+2)}\right)$} \\
\midrule
FECRP1 & 157 & 0.02 & 83.85*** & -0.12 (0.15) &  &  &  \\
FECRP2 & 157 & 0.05 & 81.43*** & -0.13 (0.15) & -4.82 (3.74) &  &  \\
FECRP3 & 157 & 0.05 & 41.30*** & -0.13 (0.15) & -5.73 (3.91) & -0.82 (0.76) &  \\
FECRP4 & 141 & 0.08 & 37.78*** & -0.17 (0.15) & -7.59 (4.02) & -1.02 (0.75) & 0.16 (0.68) \\
\midrule
\multicolumn{8}{l}{Dependent variable: $\log_{10}\left(CRP_{(c,t+3)}\right)$} \\
\midrule
FECRP1 & 139 & 0.00 & 83.20*** & -0.04 (0.14) &  &  &  \\
FECRP2 & 139 & 0.00 & 77.36*** & -0.05 (0.14) & -1.58 (4.09) &  &  \\
FECRP3 & 139 & 0.01 & 37.15*** & -0.05 (0.14) & -2.55 (4.23) & -1.02 (0.68) &  \\
FECRP4 & 126 & 0.04 & 33.03*** & -0.08 (0.15) & -2.77 (4.45) & -1.12 (0.80) & -1.00 (0.57) \\
\midrule
\multicolumn{8}{l}{Dependent variable: $\log_{10}\left(CRP_{(c,t+4)}\right)$} \\
\midrule
FECRP1 & 121 & 0.00 & 79.31*** & 0.00 (0.08) &  &  &  \\
FECRP2 & 121 & 0.00 & 72.60*** & 0.00 (0.08) & -0.83 (4.06) &  &  \\
FECRP3 & 121 & 0.00 & 33.33*** & -0.00 (0.08) & -1.30 (3.93) & -0.61 (0.89) &  \\
FECRP4 & 111 & 0.02 & 29.33*** & 0.01 (0.09) & -1.40 (3.98) & -0.72 (1.07) & -0.68 (0.34) \\
\midrule
\multicolumn{8}{l}{Dependent variable: $\log_{10}\left(CRP_{(c,t+5)}\right)$} \\
\midrule
FECRP1 & 103 & 0.01 & 74.22*** & 0.12 (0.06) &  &  &  \\
FECRP2 & 103 & 0.02 & 67.59*** & 0.12 (0.06) & -2.32 (3.92) &  &  \\
FECRP3 & 103 & 0.02 & 31.48*** & 0.12 (0.07) & -2.15 (4.14) & 0.41 (1.05) &  \\
FECRP4 & 96 & 0.03 & 25.68*** & 0.14 (0.08) & -2.88 (4.30) & 0.60 (1.05) & -0.22 (0.68) \\
\midrule
\multicolumn{8}{l}{Dependent variable: $\log_{10}\left(CRP_{(c,t+1:3)}\right)$} \\
\midrule
FECRP1 & 136 & 0.04 & 152.87*** & -0.14 (0.16) &  &  &  \\
FECRP2 & 136 & 0.09 & 153.66*** & -0.14 (0.16) & -5.13 (4.43) &  &  \\
FECRP3 & 136 & 0.11 & 77.32*** & -0.15 (0.15) & -6.15 (4.43) & -1.01 (0.69) &  \\
FECRP4 & 124 & 0.19 & 74.31*** & -0.19 (0.16) & -7.85 (4.12) & -1.13 (0.81) & -0.61 (0.50) \\
\midrule
\multicolumn{8}{l}{Dependent variable: $\log_{10}\left(CRP_{(c,t+1:5)}\right)$} \\
\midrule
FECRP1 & 96 & 0.02 & 259.31*** & -0.08 (0.13) &  &  &  \\
FECRP2 & 96 & 0.05 & 251.74*** & -0.07 (0.13) & -3.58 (4.83) &  &  \\
FECRP3 & 96 & 0.05 & 113.72*** & -0.07 (0.13) & -3.63 (4.90) & -0.07 (0.65) &  \\
FECRP4 & 90 & 0.09 & 98.08*** & -0.10 (0.13) & -4.18 (5.05) & -0.21 (0.79) & -0.55 (0.61) \\
\bottomrule
\end{tabular}  \tiny
        \caption{Fixed-effects regression results considering convictions for corruption. Columns include model ID, the number of observations $n$ used to fit each model, goodness of fit $R^2$ (within), F-statistic for poolability, and coefficient estimates (standard errors in parentheses). We use * to denote significance at the $0.05$ $p$-level; ** at the $0.01$ $p$-level, and *** at the $0.001$ $p$-level.}
        \label{tab:fixed_res_CRP}
\end{table}
\vfill
\begin{table}[p]
    \centering
    \rowcolors{2}{gray!25}{white}
    \setlength{\tabcolsep}{6pt} 
    \tiny
    \begin{tabular}{cccccccc}
\toprule
Model ID & $n$ & $R^{2}$ (within) & F (poolability) & $\log_{10}\left(STR_{(c,t)}\right)$ & $\log_{10}\left(POP_{(c,t)}\right)$ & $\log_{10}\left(SGDP_{(c,t)}\right)$ & $\log_{10}\left(POL_{(c,t)}\right)$ \\
\midrule\midrule
\multicolumn{8}{l}{Dependent variable: $\log_{10}\left(DRT_{(c,t+1)}\right)$} \\
\midrule
FEDRT1 & 124 & 0.11 & 84.09*** & 0.13* (0.06) &  &  &  \\
FEDRT2 & 124 & 0.12 & 76.88*** & 0.13* (0.06) & -1.40 (3.16) &  &  \\
FEDRT3 & 124 & 0.16 & 46.33*** & 0.15* (0.06) & -0.21 (2.83) & 1.00* (0.44) &  \\
FEDRT4 & 115 & 0.19 & 29.61*** & 0.15* (0.06) & -0.36 (2.64) & 1.05* (0.49) & -0.51 (0.58) \\
\midrule
\multicolumn{8}{l}{Dependent variable: $\log_{10}\left(DRT_{(c,t+2)}\right)$} \\
\midrule
FEDRT1 & 123 & 0.05 & 87.71*** & 0.08 (0.05) &  &  &  \\
FEDRT2 & 123 & 0.05 & 77.89*** & 0.08 (0.05) & 0.22 (3.14) &  &  \\
FEDRT3 & 123 & 0.08 & 49.02*** & 0.10 (0.05) & 1.13 (2.95) & 0.80* (0.38) &  \\
FEDRT4 & 112 & 0.14 & 36.45*** & 0.08 (0.05) & 0.94 (2.57) & 0.95* (0.43) & -0.79 (0.70) \\
\midrule
\multicolumn{8}{l}{Dependent variable: $\log_{10}\left(DRT_{(c,t+3)}\right)$} \\
\midrule
FEDRT1 & 111 & 0.02 & 89.29*** & 0.05 (0.03) &  &  &  \\
FEDRT2 & 111 & 0.03 & 79.53*** & 0.05 (0.03) & 0.99 (3.33) &  &  \\
FEDRT3 & 111 & 0.07 & 48.53*** & 0.07 (0.04) & 1.81 (3.08) & 0.96* (0.44) &  \\
FEDRT4 & 101 & 0.07 & 31.82*** & 0.06 (0.04) & 1.30 (2.90) & 0.97* (0.48) & -0.26 (0.47) \\
\midrule
\multicolumn{8}{l}{Dependent variable: $\log_{10}\left(DRT_{(c,t+4)}\right)$} \\
\midrule
FEDRT1 & 98 & 0.03 & 108.64*** & 0.06 (0.03) &  &  &  \\
FEDRT2 & 98 & 0.05 & 97.70*** & 0.06 (0.04) & 1.50 (3.18) &  &  \\
FEDRT3 & 98 & 0.07 & 55.74*** & 0.07* (0.04) & 1.88 (3.10) & 0.66 (0.53) &  \\
FEDRT4 & 89 & 0.05 & 33.50*** & 0.06* (0.03) & 1.35 (3.09) & 0.66 (0.52) & 0.09 (0.30) \\
\midrule
\multicolumn{8}{l}{Dependent variable: $\log_{10}\left(DRT_{(c,t+5)}\right)$} \\
\midrule
FEDRT1 & 85 & 0.08 & 135.58*** & 0.09* (0.04) &  &  &  \\
FEDRT2 & 85 & 0.11 & 124.77*** & 0.09* (0.04) & 2.33 (3.00) &  &  \\
FEDRT3 & 85 & 0.11 & 72.35*** & 0.09* (0.04) & 2.36 (2.95) & 0.30 (0.63) &  \\
FEDRT4 & 78 & 0.09 & 43.72*** & 0.08* (0.03) & 1.74 (3.08) & 0.30 (0.66) & 0.16 (0.33) \\
\midrule
\multicolumn{8}{l}{Dependent variable: $\log_{10}\left(DRT_{(c,t+1:3)}\right)$} \\
\midrule
FEDRT1 & 99 & 0.07 & 103.15*** & 0.09 (0.05) &  &  &  \\
FEDRT2 & 99 & 0.07 & 87.53*** & 0.09 (0.05) & -0.12 (3.58) &  &  \\
FEDRT3 & 99 & 0.11 & 49.71*** & 0.10 (0.06) & 0.86 (3.40) & 0.89** (0.32) &  \\
FEDRT4 & 93 & 0.20 & 34.65*** & 0.09 (0.06) & 0.79 (2.91) & 0.98* (0.39) & -0.86 (0.72) \\
\midrule
\multicolumn{8}{l}{Dependent variable: $\log_{10}\left(DRT_{(c,t+1:5)}\right)$} \\
\midrule
FEDRT1 & 66 & 0.03 & 137.57*** & 0.05 (0.05) &  &  &  \\
FEDRT2 & 66 & 0.03 & 106.18*** & 0.05 (0.05) & 0.64 (4.14) &  &  \\
FEDRT3 & 66 & 0.13 & 53.78*** & 0.07 (0.05) & 1.20 (4.03) & 1.05*** (0.29) &  \\
FEDRT4 & 64 & 0.28 & 38.79*** & 0.06 (0.06) & 0.65 (3.29) & 0.75* (0.31) & -1.31 (0.92) \\
\bottomrule
\end{tabular}  \tiny
    \caption{Fixed-effects regression results considering convictions for drug trafficking. Columns include model ID, the number of observations $n$ used to fit each model, goodness of fit $R^2$ (within), F-statistic for poolability, and coefficient estimates (standard errors in parentheses). We use * to denote significance at the $0.05$ $p$-level; ** at the $0.01$ $p$-level, and *** at the $0.001$ $p$-level.}
    \label{tab:fixed_res_DRT}
\end{table}

\begin{table}[p]
    \centering
    \rowcolors{2}{gray!25}{white}
    \setlength{\tabcolsep}{6pt} 
    \tiny
    \begin{tabular}{cccccccc}
\toprule
Model ID & $n$ & $R^{2}$ (within) & F (poolability) & $\log_{10}\left(STR_{(c,t)}\right)$ & $\log_{10}\left(POP_{(c,t)}\right)$ & $\log_{10}\left(SGDP_{(c,t)}\right)$ & $\log_{10}\left(POL_{(c,t)}\right)$ \\
\midrule\midrule
\multicolumn{8}{l}{Dependent variable: $\log_{10}\left(FRD_{(c,t+1)}\right)$} \\
\midrule
FEFRD1 & 181 & 0.00 & 38.49*** & -0.03 (0.13) &  &  &  \\
FEFRD2 & 181 & 0.10 & 42.06*** & -0.03 (0.12) & -6.92* (3.19) &  &  \\
FEFRD3 & 181 & 0.11 & 42.11*** & -0.04 (0.12) & -7.84** (3.00) & -0.82 (0.86) &  \\
FEFRD4 & 163 & 0.18 & 37.26*** & 0.05 (0.08) & -8.30** (2.59) & -1.07 (0.61) & -1.00 (0.72) \\
\midrule
\multicolumn{8}{l}{Dependent variable: $\log_{10}\left(FRD_{(c,t+2)}\right)$} \\
\midrule
FEFRD1 & 179 & 0.00 & 38.51*** & -0.02 (0.11) &  &  &  \\
FEFRD2 & 179 & 0.08 & 42.12*** & -0.01 (0.11) & -6.40* (2.97) &  &  \\
FEFRD3 & 179 & 0.11 & 43.29*** & -0.03 (0.10) & -7.97** (2.94) & -1.43 (0.75) &  \\
FEFRD4 & 159 & 0.18 & 35.83*** & 0.02 (0.09) & -7.85** (2.69) & -1.71** (0.64) & -1.14 (0.61) \\
\midrule
\multicolumn{8}{l}{Dependent variable: $\log_{10}\left(FRD_{(c,t+3)}\right)$} \\
\midrule
FEFRD1 & 158 & 0.00 & 37.84*** & -0.01 (0.08) &  &  &  \\
FEFRD2 & 158 & 0.04 & 39.35*** & -0.01 (0.08) & -4.73 (2.98) &  &  \\
FEFRD3 & 158 & 0.07 & 40.19*** & -0.02 (0.08) & -6.03 (3.24) & -1.28 (0.74) &  \\
FEFRD4 & 142 & 0.10 & 31.13*** & 0.00 (0.08) & -5.71 (3.09) & -1.41* (0.71) & -0.82 (0.44) \\
\midrule
\multicolumn{8}{l}{Dependent variable: $\log_{10}\left(FRD_{(c,t+4)}\right)$} \\
\midrule
FEFRD1 & 138 & 0.01 & 47.23*** & 0.05 (0.05) &  &  &  \\
FEFRD2 & 138 & 0.02 & 47.41*** & 0.05 (0.06) & -2.90 (2.37) &  &  \\
FEFRD3 & 138 & 0.04 & 47.90*** & 0.05 (0.06) & -3.71 (2.55) & -0.93 (0.71) &  \\
FEFRD4 & 126 & 0.07 & 38.85*** & 0.07 (0.06) & -3.50 (2.36) & -0.95 (0.70) & -0.65* (0.29) \\
\midrule
\multicolumn{8}{l}{Dependent variable: $\log_{10}\left(FRD_{(c,t+5)}\right)$} \\
\midrule
FEFRD1 & 120 & 0.01 & 52.25*** & 0.06 (0.06) &  &  &  \\
FEFRD2 & 120 & 0.01 & 51.20*** & 0.06 (0.06) & -1.23 (1.71) &  &  \\
FEFRD3 & 120 & 0.02 & 51.05*** & 0.06 (0.06) & -0.87 (1.89) & 0.56 (0.89) &  \\
FEFRD4 & 112 & 0.10 & 46.75*** & 0.10 (0.05) & -1.13 (1.64) & 0.51 (0.89) & -1.00*** (0.28) \\
\midrule
\multicolumn{8}{l}{Dependent variable: $\log_{10}\left(FRD_{(c,t+1:3)}\right)$} \\
\midrule
FEFRD1 & 146 & 0.00 & 49.37*** & 0.01 (0.09) &  &  &  \\
FEFRD2 & 146 & 0.07 & 52.87*** & 0.00 (0.09) & -5.52 (3.52) &  &  \\
FEFRD3 & 146 & 0.10 & 54.23*** & -0.00 (0.09) & -6.84 (3.56) & -1.24 (0.74) &  \\
FEFRD4 & 132 & 0.22 & 47.21*** & 0.02 (0.08) & -6.52* (3.00) & -1.34* (0.60) & -1.26 (0.64) \\
\midrule
\multicolumn{8}{l}{Dependent variable: $\log_{10}\left(FRD_{(c,t+1:5)}\right)$} \\
\midrule
FEFRD1 & 99 & 0.02 & 100.30*** & 0.06 (0.05) &  &  &  \\
FEFRD2 & 99 & 0.11 & 108.65*** & 0.06 (0.06) & -4.93 (3.53) &  &  \\
FEFRD3 & 99 & 0.17 & 115.32*** & 0.06 (0.05) & -5.81 (3.51) & -1.19 (0.67) &  \\
FEFRD4 & 93 & 0.34 & 117.29*** & 0.04 (0.07) & -5.30* (2.54) & -1.38* (0.64) & -1.41* (0.56) \\
\bottomrule
\end{tabular} 
    \caption{Fixed-effects regression results considering convictions for fraud. Columns include model ID, the number of observations $n$ used to fit each model, goodness of fit $R^2$ (within), F-statistic for poolability, and coefficient estimates (standard errors in parentheses). We use * to denote significance at the $0.05$ $p$-level; ** at the $0.01$ $p$-level, and *** at the $0.001$ $p$-level.}
    \label{tab:fixed_res_FRD}  
\end{table}

\newpage

\newpage

\section{Basic Poisson Model} \label{appendix:poisson}

As a final robustness check, we estimate Poisson count models using the accumulated number of money-laundering convictions over years $t+1$ to $t+3$ as the dependent variable (not the average). All models are unconditional Poisson GLMs with country indicator variables and use population as an exposure offset. Results are shown in Table \ref{tab:poisson_count_robustness}. The results are consistent with our main finding: once either (i) a linear time trend or (ii) time fixed effects are included, the STR coefficient becomes insignificant (even changing sign). We stress that the results only are intended as a further robustness check; the specification is affected by serial correlation and likely over-dispersion.

\begin{table}[H]
    \centering
    \scriptsize
    \begin{tabular}{lrrllll}
\toprule
Model & $n$ & AIC & $\log_{10}\left(STR_{(c,t)}\right)$ & $\log_{10}\left(SGDP_{(c,t)}\right)$ & $\log_{10}\left(POL_{(c,t)}\right)$ & $t$ \\
\midrule
PCFE1 & 122 & 4266.87 & 0.22 (0.33) &  &  &  \\
PCFE2 & 122 & 4238.39 & 0.14 (0.32) & -1.33 (4.16) &  &  \\
PCFE3 & 116 & 3557.84 & 0.56** (0.21) & -3.23 (3.03) & -0.53 (1.14) &  \\
PCFET1 & 122 & 2913.22 & -0.50 (0.30) &  &  & 0.09** (0.03) \\
PCFET2 & 122 & 2851.14 & -0.63 (0.35) & -1.97 (2.54) &  & 0.09*** (0.03) \\
PCFET3 & 116 & 2367.26 & -0.24 (0.27) & -3.73* (1.45) & -0.52 (0.52) & 0.08*** (0.02) \\
PCFETF1 & 122 & 2846.01 & -0.43 (0.37) &  &  &  \\
PCFETF2 & 122 & 2674.45 & -0.58 (0.45) & -3.80 (2.79) &  &  \\
PCFETF3 & 116 & 2038.50 & -0.13 (0.28) & -5.54*** (1.30) & -1.12*** (0.31) &  \\
\bottomrule
\end{tabular}

    \caption{ Poisson count-model robustness check. The dependent variable is the accumulated number of money-laundering convictions over years $t+1$ to $t+3$ (not the average). PCFET models add a linear time trend; PCFETF models add time fixed effects. Standard errors are clustered by country (reported in parentheses). We use * to denote significance at the 0.05 level, ** at the 0.01 level, and *** at the 0.001 level. }
    \label{tab:poisson_count_robustness}
\end{table}

\newpage

\printbibliography

\end{document}